# On the evolutions of induction zone structure in wedge-stabilized oblique detonation with water mist flows

Hongbo Guo[1,2], Yong Xu[2], Shuying Li[1], Huangwei Zhang[2,*]

[1] *College of Power and Energy Engineering, Harbin Engineering University*
*Harbin 150001, People's Republic of China*
[2] *Department of Mechanical Engineering, National University of Singapore*
*9 Engineering Drive 1, Singapore, 117576, Republic of Singapore*

## Abstract

Two-dimensional wedge-stabilized oblique detonations in stoichiometric and fuel-lean $H_2/O_2/Ar$ mixtures with water mists are studied with Eulerian-Lagrangian method. The effects of water droplet mass flow rate on flow and chemical structures in the induction zone, as well as physical / chemical roles of water vapor, are investigated. The results show that the oblique detonation wave (ODW) can stand in a range of water mass flow rates for both stoichiometric and fuel-lean mixtures. With increased droplet mass flow rate, the deflagration front in the induction zone is distorted and becomes zigzagged, but the transition mode from oblique shock wave (OSW) to ODW does not change. Moreover, the initiation and transition locations monotonically increase, and the OSW and ODW angles decrease, due to droplet evaporation and water vapor dilution in the induction region. For fuel-lean mixtures, the sensitivity of characteristic locations to the droplet loading variations is mild, which signifies better intrinsic stability and resilience to the oncoming water droplets. The chemical explosiveness of the gaseous mixture between the lead shock and reaction front is studied with the chemical explosive method analysis. The smooth transition is caused by the highly enhanced reactivity of the gas immediately behind the curved shock, intensified by the compression waves. Nonetheless, the abrupt transition results from the intersection between the beforehand generated detonation wave in the induction zone and OSW. Besides, the degree to which the gas chemical reactivity in the induction zone for fuel-lean mixtures is reduced by evaporating droplets is generally lower than that for stoichiometric gas. Also, physical and chemical effects of water vapor from liquid droplets result in significant differences in ODW initiation and morphology. When abrupt transition occurs, both physical and chemical effects of water vapour influence transition location and ODW angle, but only physical one is important for smooth transition.

[*]Corresponding author. Tel.: +65 6516 2557; Fax: +65 6779 1459.
*E-mail address*: huangwei.zhang@nus.edu.sg.

## 1. Introduction

Oblique detonation engine (ODE) is a promising propulsion technology for air-breathing hypersonic aircraft operating at high flight Mach number [1,2]. It not only has the advantages of scramjet [3,4], but also enjoys high thermal cycle efficiency because of oblique detonation combustion [5]. Oblique detonation wave (ODW) is a fundamental combustion mode in ODE, and results from the coupling between an oblique shock and chemical reactions of supersonic combustible mixture flowing past a wedge [6]. Under some practical conditions, unsteadiness or non-uniformity of the supersonic flows may exist. For instance, variations of the flight speed or altitude of oblique detonation engine would change inflow conditions, inducing imperfect reactant mixing and compositional stratification. Besides, liquid fuel vapor concentration may not be uniform if the liquid fuel is not well atomized, resulting in differentiated evaporation rates. How an ODW is initiated and stabilized under such realistic scenarios are still not well understood.

Most of the previous studies about ODW initiation and stabilization are focused on idealized (e.g., premixed, uniform, or homogeneous) approaching combustible mixtures. For instance, Pratt et al. [7] derive analytical solutions between oblique shock/detonation wave angle and inflow conditions through thermodynamic analysis. Li et al. [6] find that oblique detonation is composed of a nonreactive oblique shock, an induction zone followed by a reaction zone, and an oblique detonation. Figueira da Silva et al. [8] reveal various regions behind the oblique shock wave (OSW) depending on local chemical and gas dynamics properties. Verreault et al. [9] investigate the dynamics of transverse waves in oblique detonations and find that spatial oscillations from the ODW transition evolve into transverse waves propagating downstream. Moreover, two OSW-ODW transition modes are identified [10,11]: abrupt transition connected by a multi-wave point, and smooth transition featured by a curved shock. Recent studies by Teng et al. [12] observe more complex wave systems, and different wave structures in the transition zone are also discussed.

How the induction zone and oblique detonation wave evolve subject to non-uniformity or perturbation of thermochemical state in the combustible mixture is also studied. For instance, Li et al. [6] study how the induction zone changes when a density perturbation is introduced, and the oblique detonation structure is shown to have good resilience. Similar resilience is also studied by Fusina et al. [13], who introduce some small pockets without fuel in the inflow mixtures. Besides, Iwata et al. [14] simulate oblique detonations with inhomogeneous mixtures, and their results indicate that the deflagration fronts can be "V" or "V+Y"-shaped. The distorted deflagration front is also observed by Guo et al. [15] when non-uniform equivalence ratios are considered. Fang et al. [16] find that ODW can remain stable, even at high equivalence ratio gradient. Recently, Ren et al. [17] investigate the response of ODW to a (time-evolving) sinusoidal variation of inflow equivalence ratio. The results show that for low-to-medium fluctuations the increasing amplitude only affects the ODW initiation and leads to ODW surface instability, and the ODW structures become more robust in fuel-lean situations.

It is well known that liquid fuels have numerous advantages, e.g., high energy density, easy storage, and possibility to be used as engine coolant. Therefore, they have good potential for detonation engine applications. Indeed, feasibility to utilize liquid fuels for laboratory scale detonation propulsion test rigs has been confirmed through deliberately designed combustor and fuel injector. For instance, Fan et al. [18] successfully conduct two-phase pulse detonation engine experiments with liquid $C_8H_{16}$/air mixture. Besides, Kindracki et al. [19] investigate the influence of liquid kerosene properties on initiation and propagation of rotating detonations. Nonetheless, use of liquid fuels may lead to highly inhomogeneous mixtures and investigations on ODW with liquid fuels are rather limited in the open literature. Recently, Ren et al. [20,21] study the effects of evaporating droplets and spray equivalence ratios on ODW initiation in two-phase kerosene-air mixtures. The results show that the variations of initiation length and transition pressure are dominated by evaporative cooling and heat release, respectively. However, they have not studied how the OSW-ODW transition mode and induction zone structure change when the gas-liquid two-phase approaching flows are considered.

Typically, under supersonic flow conditions, liquid fuels need to be well sprayed for efficient evaporation and vapor combustion due to short residence time in the chamber. This leads to a large number of dispersed fine droplets in the incoming flows. These droplets would have strong and multi-faceted coupling with the continuous gas phase. For example, droplet evaporation leads to latent heat absorption and vapor addition, thereby modulating the thermochemical state of the mixture and combustion morphology [22]. The objective of this work is to study how the evaporating droplets affect the initiation mode and induction zone structure of the oblique detonation induced by a wedge. Eulerian-Lagrangian method is used to simulate the multi-species compressible gas-liquid two-phase flows. Hydrogen will be considered as the fuel, and the disperse phase is ultrafine water droplets. A parametric analysis is performed to reveal the effects of water droplet mass flow rate on detonation initiation and induction zone structure in both abrupt and smooth transitions from shock to detonation.

## 2. Mathematical model and computational method

The Eulerian-Lagrangian method is employed to simulate wedge-induced hydrogen detonations in water fog environment. The Eulerian gas and Lagrangian liquid droplet governing equations are solved by a compressible two-phase reacting flow solver, *RYrhoCentralFoam* [23,24]. It is developed from a fully compressible flow solver *rhoCentralFoam* in OpenFOAM 6.0 [25]. The details about the equations and models are presented below.

### 2.1 Gas phase

The equations of mass, momentum, energy, and species mass fraction are solved for multi-species, two-phase, reactive, compressible flows. They respectively read

$$\frac{\partial \rho}{\partial t} + \nabla \cdot [\rho \mathbf{u}] = S_{mass}, \quad (1)$$

$$\frac{\partial (\rho \mathbf{u})}{\partial t} + \nabla \cdot [\mathbf{u}(\rho \mathbf{u})] + \nabla p + \nabla \cdot \mathbf{T} = \mathbf{S}_{mom}, \quad (2)$$

$$\frac{\partial (\rho E)}{\partial t} + \nabla \cdot [\mathbf{u}(\rho E + p)] + \nabla \cdot [\mathbf{T} \cdot \mathbf{u}] + \nabla \cdot \mathbf{j} = \dot{\omega}_T + S_{energy}, \quad (3)$$

$$\frac{\partial (\rho Y_m)}{\partial t} + \nabla \cdot [\mathbf{u}(\rho Y_m)] + \nabla \cdot \mathbf{s_m} = \dot{\omega}_m + S_{species,m}, (m = 1, \dots M-1). \quad (4)$$

In above equations, $t$ is time and $\nabla \cdot (\cdot)$ is the divergence operator. $\rho$ is the gas density, $\mathbf{u}$ is the velocity vector, and $p$ is the pressure updated from the equation of state, i.e., $p = \rho RT$. $T$ is the gas temperature and $R$ is specific gas constant, calculated from $R = R_u \sum_{m=1}^{M} Y_m W_m^{-1}$. $W_m$ is the molar weight of $m$-th species and $R_u = 8.314$ J/(mol·K) is the universal gas constant. $Y_m$ is the mass fraction of $m$-th species and $M$ is the total species number. $E \equiv e_s + |\mathbf{u}|^2/2$ is the total non-chemical energy, in which $e_s = h_s - p/\rho$ is the sensible internal energy and $h_s$ is the sensible enthalpy [26].

The viscous stress tensor $\mathbf{T}$ in Eq. (2) is modelled by $\mathbf{T} = -2\mu[\mathbf{D} - \mathrm{tr}(\mathbf{D})\mathbf{I}/3]$. $\mu$ is the dynamic viscosity, following the Sutherland's law. $\mathbf{D} \equiv [\nabla\mathbf{u} + (\nabla\mathbf{u})^T]/2$ is the deformation gradient tensor and $\mathbf{I}$ is the unit tensor. In addition, $\mathbf{j}$ in Eq. (3) is the diffusive heat flux, modelled with Fourier's law, i.e., $\mathbf{j} = -k\nabla T$. Thermal conductivity $k$ is calculated using the Eucken approximation [27], i.e., $k = \mu C_v(1.32 + 1.37R/C_v)$. $C_v$ is the heat capacity at constant volume and derived from $C_v = C_p - R$. Here $C_p = \sum_{m=1}^{M} Y_m C_{p,m}$ is the heat capacity at constant pressure, and $C_{p,m}$ is the heat capacity at constant pressure of $m$-th species, estimated from JANAF polynomials [28].

In Eq. (4), $\mathbf{s_m} = -D_m\nabla(\rho Y_m)$ is the species mass flux. $D_m$ is calculated through $D_m = k/\rho C_p$ with a unity Lewis number assumption. Moreover, $\dot{\omega}_m$ is the production or consumption rate of $m$-th species by all $N$ reactions, and can be calculated from the reaction rate of each reaction, i.e., $\dot{\omega}_m = W_m \sum_{j=1}^{N} \omega_{m,j}^o$. Also, the term $\dot{\omega}_T$ in Eq. (3) represents the heat release rate from chemical reactions and is estimated as $\dot{\omega}_T = -\sum_{m=1}^{M} \dot{\omega}_m \Delta h_{f,m}^o$. $\Delta h_{f,m}^o$ is the formation enthalpy of $m$-th species.

## 2.2 Liquid phase

The Lagrangian method is used to model the dispersed liquid phase, which is composed of a large number of spherical droplets [29]. Droplet collisions are neglected because we only study dilute water sprays with initial volume fraction less than 0.1% [30]. Droplet breakup is not considered in this work. Besides, since the ratio of gas density to the water droplet material density is well below one, Basset force, history force and gravity force are neglected [30]. The temperature inside the droplet is assumed to be uniform, considering the small Biot number of water droplets. Therefore, the governing equations of mass, momentum and energy for a single droplet are

$$\frac{dm_d}{dt} = -\dot{m}_d, \tag{5}$$

$$\frac{d\mathbf{u}_d}{dt} = \frac{\mathbf{F}_d + \mathbf{F}_p}{m_d}, \quad (6)$$

$$c_{p,d}\frac{dT_d}{dt} = \frac{\dot{Q}_c + \dot{Q}_{lat}}{m_d}, \quad (7)$$

where $m_d = \pi\rho_d d_d^3/6$ is the mass of a single droplet, $\rho_d$ and $d_d$ are the droplet material density and diameter, respectively. $\mathbf{u}_d$ is the droplet velocity vector, $\mathbf{F}_d$ and $\mathbf{F}_p$ are the drag and pressure gradient force exerting on the droplet. $c_{p,d}$ is the droplet heat capacity at constant pressure, and $T_d$ is the droplet temperature.

The droplet evaporation rate, $\dot{m}_d$, in Eq. (5) is modelled through [31]

$$\dot{m}_d = \pi d_d Sh D_{ab} \rho_s \ln{(1 + X_r)}, \quad (8)$$

where the vapor molar ratio $X_r$ is estimated from

$$X_r = \frac{X_S - X_C}{1 - X_S}. \quad (9)$$

Here $X_C$ is the vapor mole fraction in the surrounding gas, and $X_S$ is the water vapor mole fraction at the droplet surface, calculated using Raoult's Law, i.e.,

$$X_S = X_m \frac{p_{sat}}{p}. \quad (10)$$

$X_m$ is the mole fraction of the condensed species in the liquid phase. $p_{sat}$ is the saturation pressure and is a function of droplet temperature $T_d$ [32], i.e.,

$$p_{sat} = p \cdot exp\left(c_1 + \frac{c_2}{T_d} + c_3 lnT_d + c_4 T_d^{c_5}\right), \quad (11)$$

where the constants $c_i$ can be found from Ref. [32]. Moreover, in Eq. (8), $\rho_S = p_S W_m / RT_S$ is the vapor density at the droplet surface, where $p_S$ is the surface vapor pressure and $T_s = (T + 2T_d)/3$ is the droplet surface temperature. $D_{ab}$ is the vapor diffusivity in the gaseous mixture [33], i.e.,

$$D_{ab} = 3.6059 \times 10^{-3} \times (1.8T_s)^{1.75} \times \frac{\alpha_l}{p_S \beta_l}, \quad (12)$$

where $\alpha_l$ and $\beta_l$ are the model constants from Ref. [34].

The Sherwood number in Eq. (8) is modelled as

$$Sh = 2.0 + 0.6 Re_d^{1/2} Sc^{1/3}, \quad (13)$$

where *Sc* is the Schmidt number of gas phase. The droplet Reynolds number in Eq. (13), $Re_d$ is defined based on the slip velocity

$$Re_d \equiv \frac{\rho d_d |\mathbf{u}_d - \mathbf{u}|}{\mu}. \tag{14}$$

The drag force $\mathbf{F}_d$ in Eq. (6) is modelled as [35]

$$\mathbf{F}_d = \frac{18\mu}{\rho_d d_d^2} \frac{C_d Re_d}{24} m_d (\mathbf{u} - \mathbf{u}_d). \tag{15}$$

The drag coefficient, $C_d$, is estimated as [35]

$$C_d = \begin{cases} 0.424, & Re_d \geq 1000, \\ \frac{24}{Re_d}\left(1 + \frac{1}{6} Re_d^{\frac{2}{3}}\right), & Re_d < 1000. \end{cases} \tag{16}$$

Besides, the pressure gradient force $\mathbf{F}_p$ in Eq. (6) is

$$\mathbf{F}_p = -V_d \nabla p. \tag{17}$$

Here $V_d$ is the volume of a single water droplet.

The convective heat transfer rate $\dot{Q}_c$ in Eq. (7) is calculated from

$$\dot{Q}_c = h_c A_d (T - T_d), \tag{18}$$

where $h_c$ is the convective heat transfer coefficient, and computed using the correlation by Ranz and Marshall [36], i.e.,

$$Nu = h_c \frac{d_d}{k} = 2.0 + 0.6 Re_d^{1/2} Pr^{1/3}, \tag{19}$$

where *Nu* and *Pr* are the Nusselt and Prandtl numbers of gas phase, respectively. In addition, the heat transfer associated with droplet evaporation, $\dot{Q}_{lat}$ in Eq. (7), is

$$\dot{Q}_{lat} = -\dot{m}_d h(T_d), \tag{20}$$

where $h(T_d)$ is the latent heat of vaporization at the droplet temperature $T_d$.

Two-way coupling between the gas and liquid phases is implemented through Particle-source-in-cell (PSI-CELL) approach [37]. The terms, $S_{mass}$, $\mathbf{S}_{mom}$, $S_{energy}$ and $S_{species,m}$ in Eqs. (1)–(4), are calculated are based on all droplets in a CFD cell, i.e.,

$$S_{mass} = \frac{1}{V_c} \sum_{i=1}^{N_d} \dot{m}_{d,i}, \tag{21}$$

$$\boldsymbol{S}_{mom} = -\frac{1}{V_c} \sum_{i=1}^{N_d} \left(-\dot{m}_{d,i} \mathbf{u}_{d,i} + \mathbf{F}_{d,i}\right), \tag{22}$$

$$S_{energy} = -\frac{1}{V_c}\sum_{i=1}^{N_d}\left(-\dot{m}_{d,i}h_g + \dot{Q}_{c,i}\right), \quad (23)$$

$$S_{species,m} = \begin{cases} S_{mass} & for \quad H_2O \quad species \\ 0 & for \quad other \quad species \end{cases}. \quad (24)$$

Here $V_c$ is the CFD cell volume, $N_d$ is the droplet number in one cell, and $h_g$ the water vapor enthalpy at the droplet temperature. Due to the dilute and fine sprays considered, the hydrodynamic force work by the droplet phase is neglected in Eq. (23).

### 2.3 Computational method

For the gas phase, cell-centered fine volume method is used to discretizing Eqs. (1)−(4). Second-order implicit backward method is employed for temporal discretization and the time step is about $5\times10^{-11}$ s. The MUSCL-type Riemann-solver-free scheme by Kurganov et al. [38] is used for the convective terms in the momentum equations, whilst the total variation diminishing scheme for the convective terms in energy and species equations. Also, second-order central differencing scheme is applied for the diffusion terms in Eqs. (2)−(4). Chemistry integration is performed with a Euler implicit method, and its accuracy has been confirmed with other ordinary differential equation solvers in our recent work [24]. A hydrogen mechanism with 9 species and 19 reactions [39] is used, which is validated against measured ignition delay and detonation cell size [24].

For the liquid phase, the water droplets are tracked based on their barycentric coordinates. The equations, i.e., Eqs. (5)−(7), are solved by first-order Euler method and the right-hand terms (e.g., $\dot{m}_d$ in Eq. 5) are integrated in a semi-implicit approach. Meanwhile, the gas properties at the droplet location are interpolated from gas phase simulation results.

The above numerical methods in the *RYrhoCentralFoam* solver have been extensively validated and verified against the experimental or theoretical data in different detonation and shock benchmarking problems [24]. It has been successfully applied for modelling various detonation and supersonic combustion problems [4043]. In particular, the accuracy of *RYrhoCentralFoam* in ODW simulations is also verified and we find that it is comparable with that of the results by Tian et al. [44]

(see details in supplementary document).

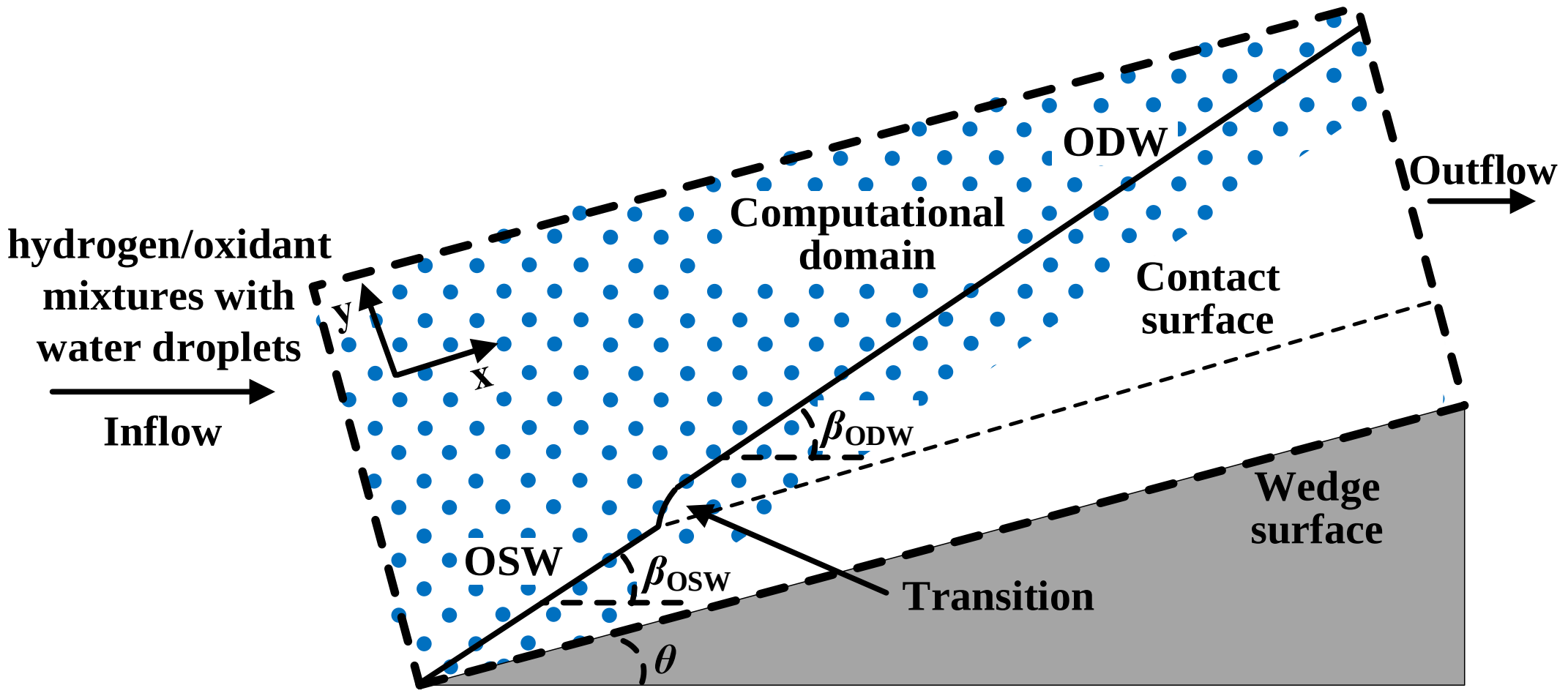

Figure 1: Schematic of a wedge-induced ODW. Blue dots: water droplets.

## 3. Physical model and numerical implementation

Oblique detonations induced by a semi-infinite wedge with an inflow of gas-liquid two-phase mixtures are studied in this work. A two-dimensional configuration is considered, as illustrated in Fig. 1. The simulated domain is enclosed by the dashed lines. The $x$ ($y$) axis is parallel (normal) to the wedge surface, whilst the origin lies at the leading edge of the wedge. Note that the length ($x$-direction) of the domain is adjusted in different simulations, to ensure that oblique detonation morphology of interest is fully captured. The wedge angle is $\theta = 25°$ in our simulations.

Dirichlet conditions are enforced for gas properties (e.g., species mass fractions, velocity, temperature, and pressure) at the left and top boundaries in Fig. 1. The gas composition of the incoming mixtures is hydrogen and oxygen with 70% (by vol.) argon dilution. Two equivalence ratios are studied, i.e., $\phi = 1.0$ and 0.5. Therefore, the species mole ratio follows $H_2/O_2/Ar = 2\phi/1/7$. Based on our simulations, they respectively correspond to abrupt and smooth transition from OSW to ODW [6,45], and therefore it is favorable to examine how these two modes respond to the dispersed droplets. The incoming flow velocity is 2447 m/s (e.g., inflow gas Mach number is 7.0 for $\phi = 1.0$, whereas 7.3 for $\phi = 0.5$), whereas the temperature and pressure are 298 K and 1.0 atm, respectively. Besides, the outlet is non-reflective, whilst non-slip and adiabatic conditions are employed for the wedge surface. A rebound condition is applied for droplet-wall interactions, in which the normal component of droplet velocities is reversed after wall collision.

Ultrafine water droplets are loaded in the incoming flows, to mimic a highly humid or foggy atmosphere. They enter the domain from left and top boundaries, as marked in Fig. 1. The droplets are assumed to be spherical, with initial diameter of $d_d^0$ = 1 μm. Such small diameter is reasonable, because the incoming flows are compressed by the preceding shocks in the inlet [1,46,47] and the droplets may have been aerodynamically fragmented. The droplet initial material density, heat capacity, and temperature are 997 kg/m$^3$, 4187 J/kg/K, and 298 K, respectively. The droplet injection velocity is assumed to be identical to that of the local carrier gas (i.e., initial slip velocity is zero). In our simulations, we consider the following droplet mass flow rates, i.e., $f_d$ = 0.208, 0.416, 0.624, 0.832 and 1.041 g/s.

The computational domain in Fig. 1 is discretized with Cartersian cells. The background mesh size is 16×16 μm$^2$, based on which two additional levels of mesh refinement are made with OpenFOAM *refineMesh* utility near the OSW and ODW, as well as the post-OSW and -ODW areas. This leads to a cell size of 4×4 μm$^2$ in the foregoing locations to sufficiently resolve the significant characteristics of OSW and ODW. The half-reaction lengths in Chapman-Jouguet detonations with $\phi$ = 0.5 and 1.0 are about 174 and 69 μm, respectively, calculated with Shock and Detonation Toolbox [48]. As such, they correspond to about 44 and 17 cells in the half-reaction zone. This resolution is comparable with or even finer than those for gaseous ODW simulations [9,4951]. Considering the droplet evaporation and heat absorption from the gas phase, more cells can be expected due to the thickened half-reaction zone in spray detonations [52,53]. Mesh sensitivity analysis (presented in supplementary document) has shown that further refining the mesh does not change the results of the present ODW simulations.

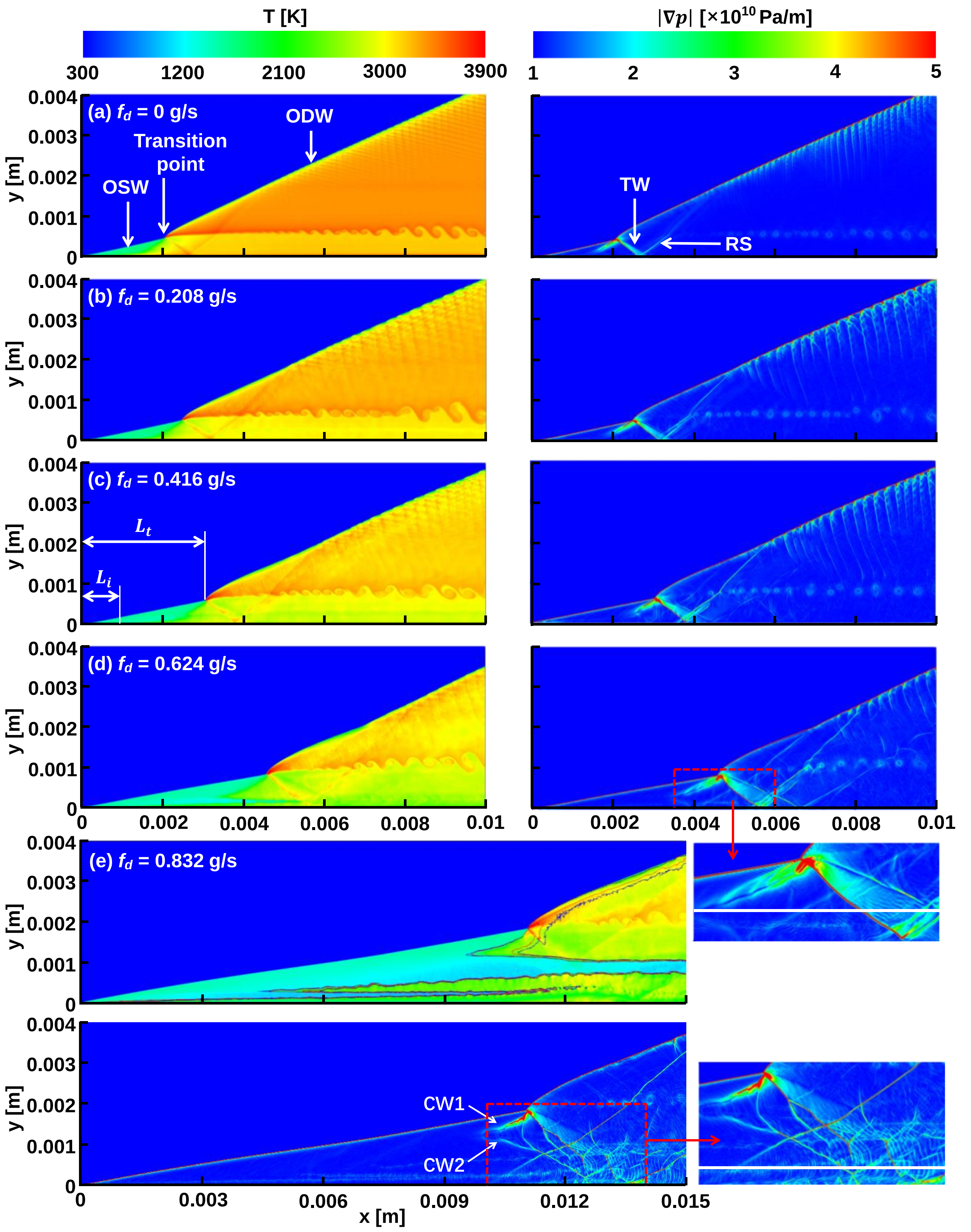

Figure 2: Gas temperature and pressure gradient magnitude in stoichiometric $H_2/O_2/Ar$ mixtures with different droplet mass flow rates. Solid lines in Fig. 2(e): isolines of heat release rate from $3\times10^{12}$ to $5\times10^{13}$ J/(m$^3$s). CW1 and CW2: compression wave, TW: transverse wave, RS: reflected shock.

## 4. Results

### 4.1 Stoichiometric $H_2/O_2/Ar$ mixture

#### 4.1.1 ODW characteristics

Oblique detonations in stoichiometric $H_2/O_2/Ar$ mixtures with different water mass flow rates (i.e., $f_d$ = 0.208–0.832 g/s) are shown in Fig. 2, about the gas temperature and pressure gradient magnitude. Generally, the ODW morphology is considerably influenced by the dispersed water droplets, although the ODWs can stably stand in the supersonic two-phase flows in all cases. However, compared with the gas-only results in Fig. 2(a), the ODW surface tends to be more cellular, and the distance between the transverse waves slightly increases as the droplet mass flow rate increases, as evidenced from the pressure gradient distributions, which indicates more pronounced instability of the oblique detonative waves [9,49]. Evident from Fig. 2(a) is an abrupt OSW-ODW transition in gas-only mixture, with a salient multi-wave point (annotated as "transition point"). This is also predicted by Teng et al. [54], under the same fuel and inflow conditions. In Figs. 2(b)−2(e), when the water droplets are loaded, this mode remains, but the transition location recedes when $f_d$ gradually increases., which will be further discussed in Section 5.1.

A closer inspection of Fig. 2 also reveals a fundamental change of the flow structure in the induction zone when water droplets are loaded. When $f_d \geq 0.624$ g/s in Figs. 2(d) and 2(e), C-shaped compression / pressure waves with upper and lower branches can be found close to the transition point (see the inset of Fig. 2d). The upper compression waves interact with the incoming droplets, and their leading edges are distorted. The lower pressure wave is still weak with $f_d$ = 0.624 g/s but become stronger when $f_d$ is further increased to 0.832 g/s, which is marked as CW2 in Fig. 2(e). In general, the intensity of CW1 is much higher than that of CW2, as can be found from the inset of Fig. 2(e). The C-shaped structure is shown to regularly oscillate with moving transition point, which results in a dynamic stabilization of the ODW. This can be clearly seen from the animation submitted with this manuscript. As $f_d$ increases, the wave structure underneath the transition becomes more complicated, characterized by the interactions between shock waves / shocklets and compression waves.

The distributions of post-OSW gas temperature and pressure with various $f_d$ are shown in Fig. 3, which are extracted along $y$ = 0.3 mm. As $f_d$ increases, the OSW gradually recedes with larger $x$ coordinates, whilst the temperature behind the OSW decreases. The latter can also be seen from the temperature contours in Fig. 2, which is caused by the heat absorption for droplet heating and gasification. Moreover, the pressure experiences a jump across oblique shock waves. Various pressure peaks are observable behind the OSW. The pressure peaks (e.g., $f_d$ = 0 and 0.208 g/s) corresponds to the position of characteristic wave structures behind the OSW, such as compression waves, transverse shock wave and reflected shock. The reader should be reminded that with increased droplet mass flow rates, the position of multi-wave point moves downstream and off the wedge surface. Therefore, the position of $y$ = 0.3 mm actually corresponds to different part of the post-OSW structure, as marked by the white lines in the insets of Fig. 2.

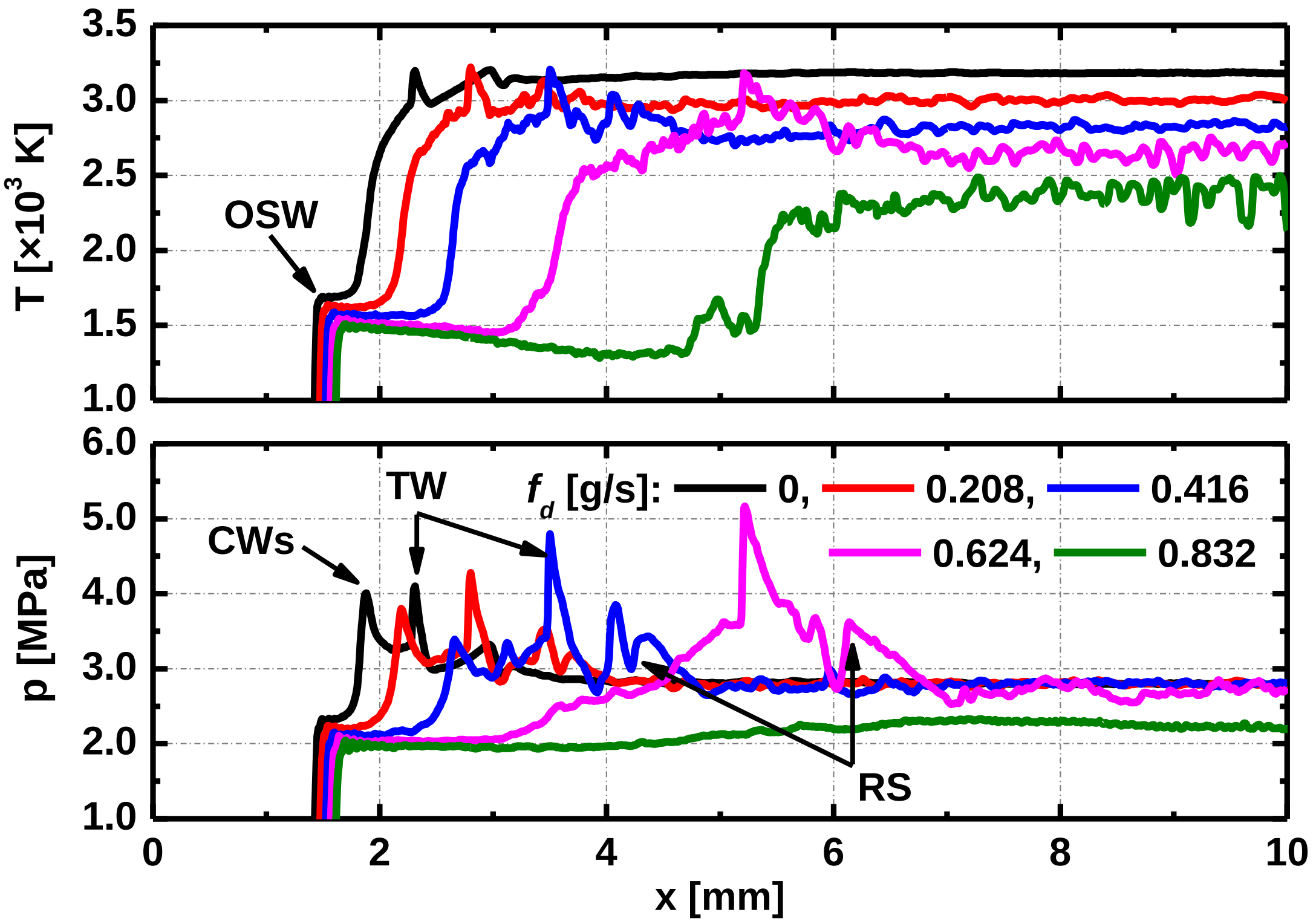

Figure 3: Profiles of gas temperature (upper) and pressure (lower) with different droplet mass flow rates. Results from $y$ = 0.3 mm. CWs: compression waves, TW: transverse wave, RS: reflected shock.

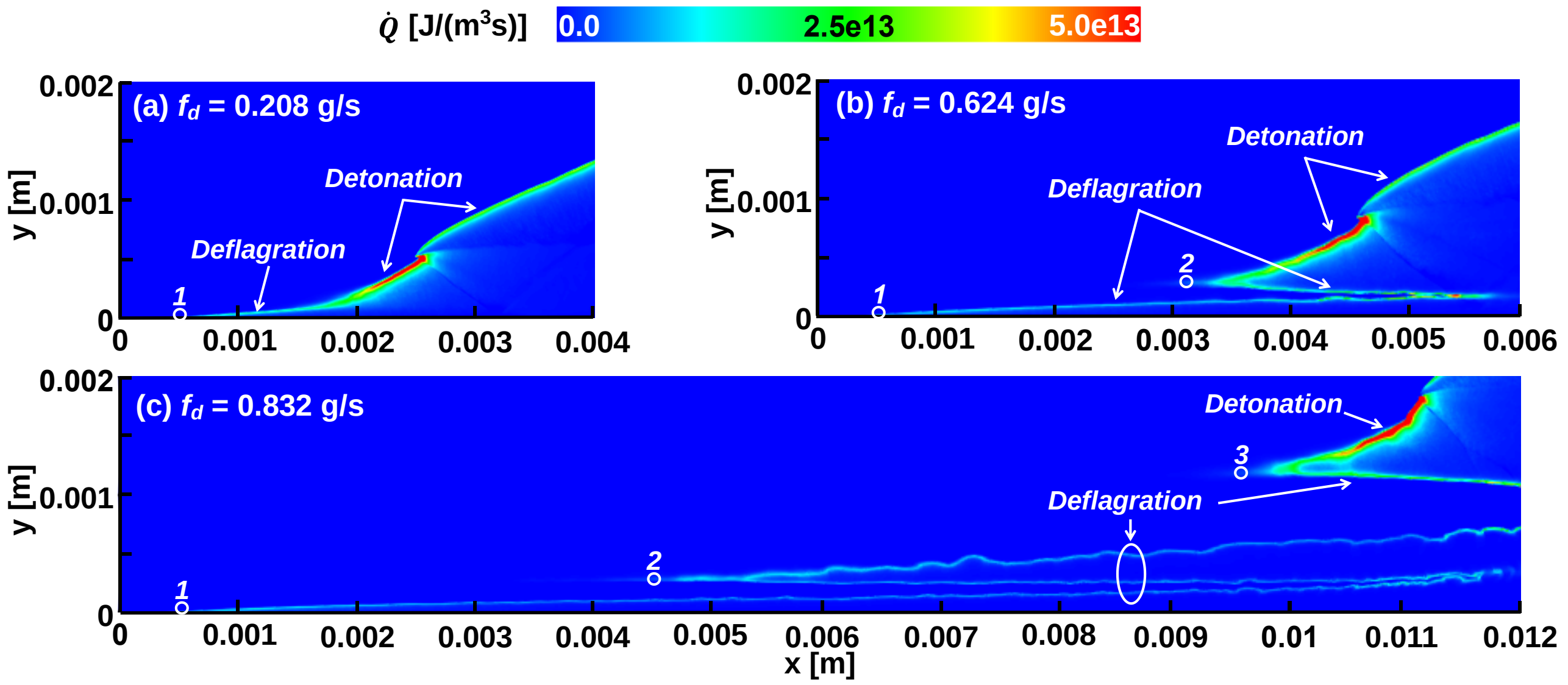

Figure 4: Distributions of combustion heat release rate with droplet mass flow rate of (a) 0.208, (b) 0.624, and (c) 0.832 g/s. Numered circle: deflagration initiation location.

Figure 4 shows the distributions of the reactive fronts with different water droplet mass flow rates. The reaction front can be identified through high combustion heat release rate (HRR). When $f_d$ = 0.208 g/s in Fig. 4(a), the deflagrative combustion is initiated at location #1 (about $x$ = 0.39 mm). At around $x$ = 2 mm, detonative combustion is initiated due to the mutual enhancement of the local heat release and shock wave, with HRR larger than $5\times10^{13}$ J/m$^3$s. This reaction front distribution is similar to that in gas-only mixture (not shown here) corresponding to Fig. 2(a). When $f_d$ is further increased to 0.624 g/s, bifurcation of the reaction front appears, with multiple initiation loci and segmented reaction fronts. Figure 4(b) shows that besides #1, a new initiation location #2 is present at around $x$ = 3 mm. The upper and lower branches of the C-shaped reaction fronts starting from #2 correspond to detonation and deflagration waves, respectively. Similar C-shaped reaction fronts are also observed in gaseous ODW with stratified inflowing premixtures [14,16] and are attributed to the non-uniform mixture composition near the wedge surface. Besides, the deflagrative fronts extending from #1 and #2 intersect at around $x$ = 6 mm, resulting in a spatially delayed deflagration fronts between #1 and #2, as compared to the reaction front morphology in Fig. 4(a) and gas-only mixture.

In Fig. 4(c), when $f_d$ = 0.832 g/s, the distributions of reaction fronts in the induction zone become more complicated. They become nominally zigzagged, with three reaction initiation points, i.e., #1, #2, and #3. The deflagrative front interaction position extending from the locations #1 and #2 is more downstream, at about $x$ = 12 mm. The distance between the deflagrative fronts from different initiation locations increases, and the heat release rate of combustion waves far away from the main multi-wave point decreases. These peculiar distributions of reaction fronts in Fig. 4 are caused by the dispersed droplets in the shocked gas, which will be examined through the droplet distribution in Section 4.1.2.

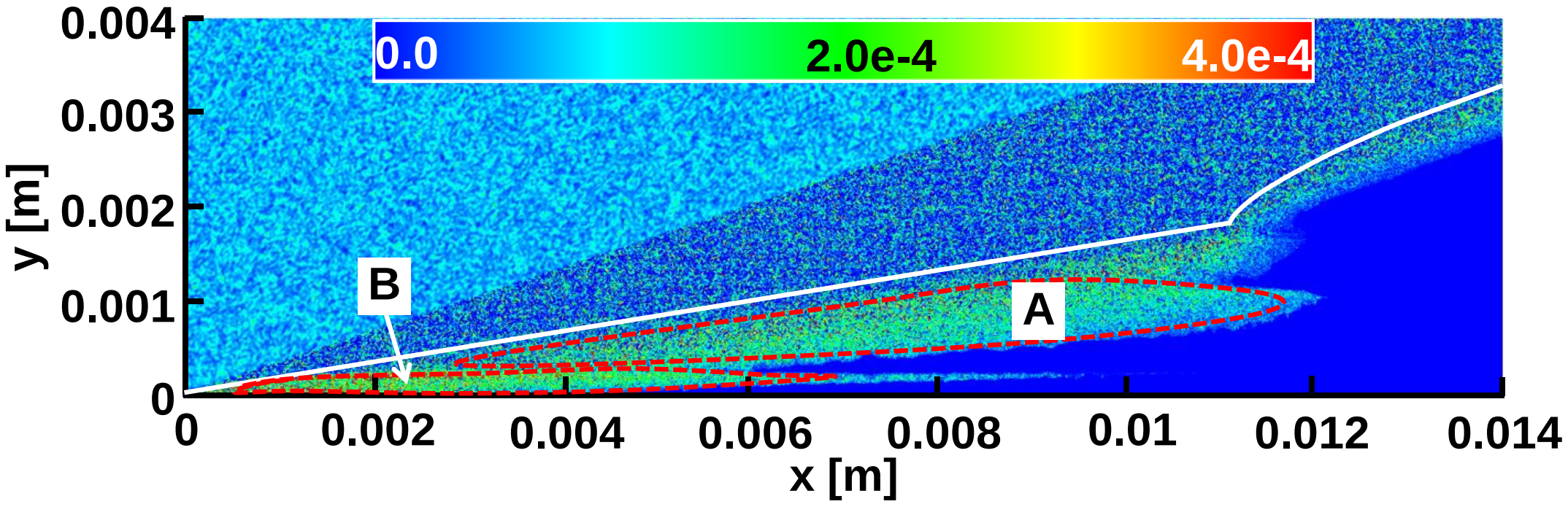

Figure 5: Distribution of droplet volume fraction. $f_d$ = 0.832 g/s. Solid line: OSW and ODW.

### 4.1.2 Droplet distribution

Figure 5 shows the distribution of droplet volume fraction when $f_d$ = 0.832 g/s. It is calcualed from the ratio of total droplet volume to gas volume in each CFD cell. Apparently, behind the OSW, there are two distinct zones (i.e., A and B) with higher volume fractions and hence larger droplet concentrations. Why the droplets are accummulated there will be interpreted in Figs. 6 and 7. Figures 6(a) – 6(d) respectively shows the Lagrangian droplets colored by their velocity, temperature, and diameter. Note that the colorless areas indicate that the droplets have already vaporized there. It can be seen from Figs. 6(a) and 6(b) that both $x$- and $y$-velocities are generally reduced behind both OSW and ODW. This is because the local flow decelerates after the shocks, and ultrafine droplets (1 μm in this study) can respond to the gas speed variations quickly due to their small Stokes numbers. However, behind the OSW, the $y$-direction velocities are greater than zero (50–100 m/s), and these droplet streams move along the narrow “channel” between the two deflagration fronts extending from #2 and #3, respectively. This leads to a local acumulation of the water droplets in zone A, indicated in Fig. 5. One can see from Fig. 6(c) that the temperatures of these droplets have been elevated to above 400 K (close to the local saturation temperature), due to the short thermal response time of fine droplets. After a finite distance behind OSW, droplet evaporation starts. This can be featured by reduced droplet size in Fig. 6(d) and pronounced mass transfer $\Delta m$ (i.e., water vapor) from the liquid phase to gas phase in Fig. 6(e). Significant energy absoprtion ($\Delta E$ < 0) from the gas phase can also be observed in Fig. 6(f). These mass and energy transfer with water mists significantly delay gas chemistry initiation, which hence leads to bifurcation between the leading points #2 and #3 marked in Fig. 4, and distorts the reaction fronts (see the HRR isolines in Fig. 6).

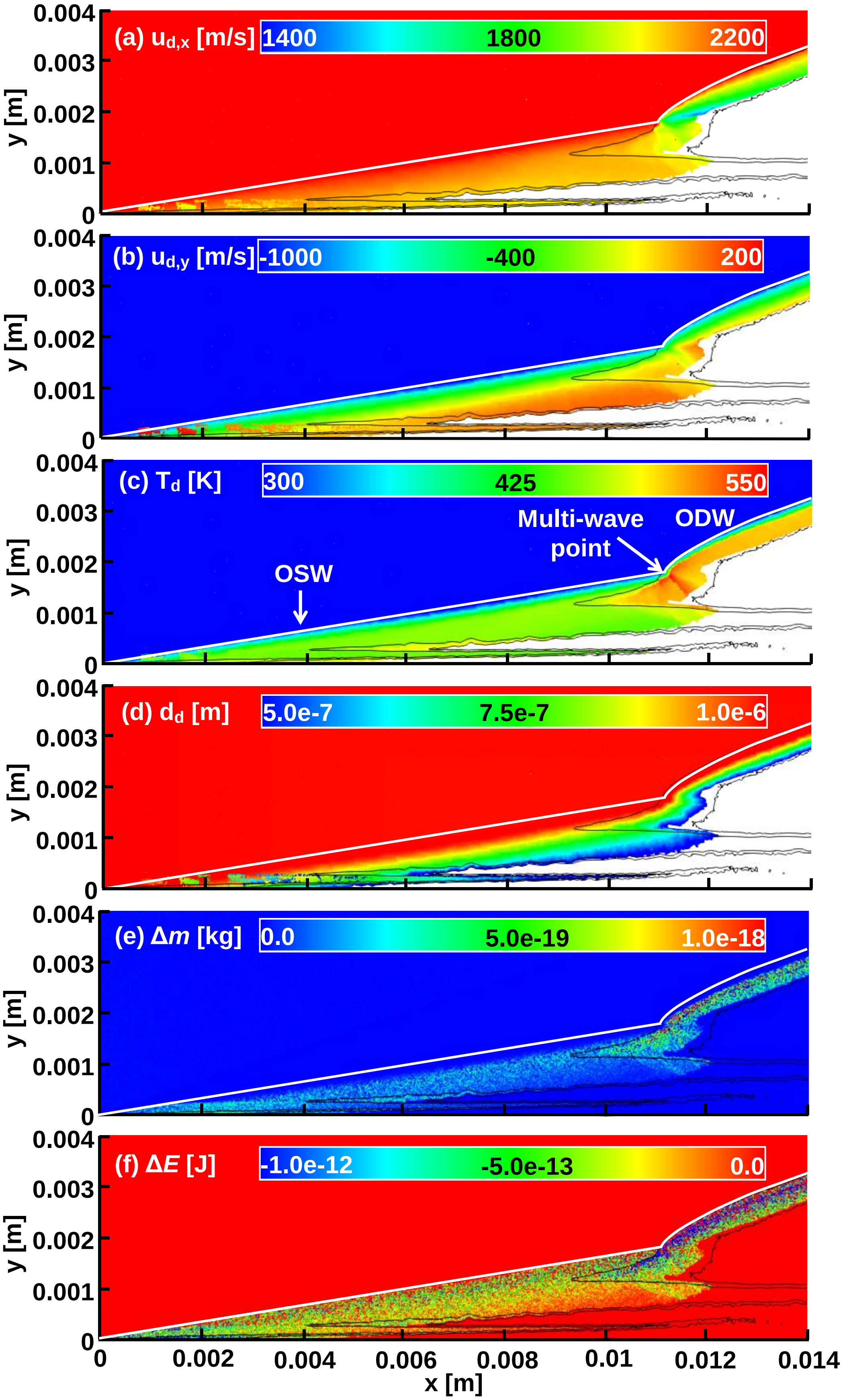

Figure 6: Lagrangian droplets colored by (a) $x$-velocity, (b) $y$-velocity, (c) temperature, (d) diameter, (e) mass transfer, and (f) energy transfer. $f_d$ = 0.832 g/s. Solid line: heat release rate of $10^{12}$ J/(m$^3$s).

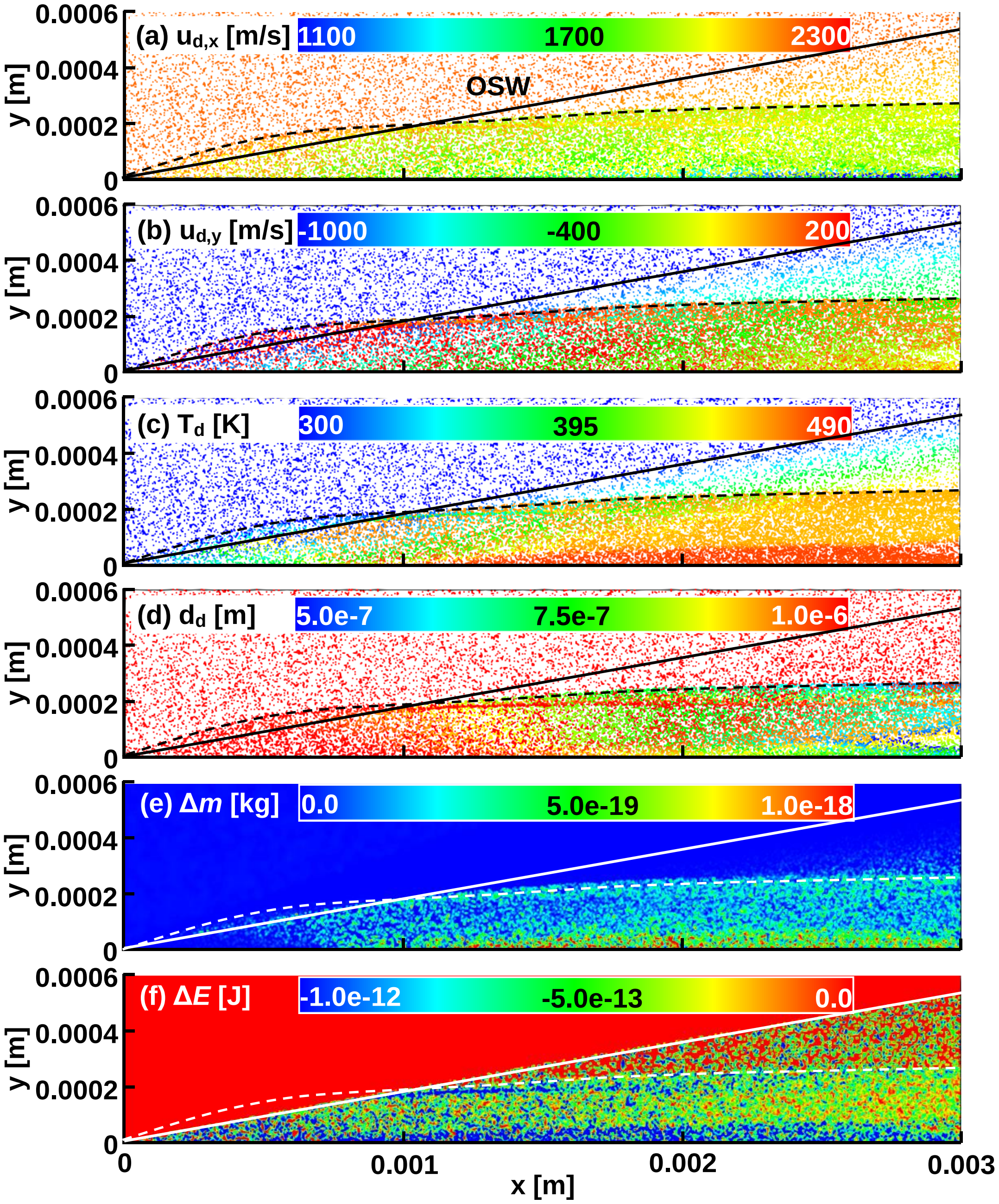

Figure 7: Close-up view of the results in Fig. 6 near the leading edge. $f_d$ = 0.832 g/s. Solid line: OSW; dashed line: approximated boundary of zone B.

To elucidate the formation of zone B in Fig. 5, Fig. 7 shows the close-up view near the leading edge. The oncoming droplets bounce up on the wedge and hence have an positive $y$-velocity, see Fig. 7(b) or Fig. 6(b). This results in an area with high water concentration, corresponding to zone B. Below the dashed line, the droplets have higher temperature and reduced diameter, because they are heated by the hot boundary layer and thereby strongly vaporize. This can be also corroborated by considerable mass and energy transfer between the droplet and gas phase, as illustrated in Figs. 7(e) and 7(f). Therefore, it has been shown that heating and evaporation of high-concentration water droplets may change the local thermochemical states in zones A and B, thereby bifurcating the reaction fronts. Similar observations are also made for other cases when reaction front bifurcation is present, e.g., $f_d$ = 0.624 g/s (due to relatively low droplet addition, only zone B is seen). How the chemical reactions modulated by the water mists in zones A and B will be quantified with chemical explosive mode analysis in Section 5.2.

## 4.2 Lean $H_2/O_2/Ar$ mixture

### 4.2.1 ODW characteristics

Fuel-lean $H_2/O_2/Ar$ mixture (ϕ = 0.5) with different water mass flow rates is studied in this section. The distributions of gas temperature and pressure gradient magnitude are shown in Fig. 8, and the evolutions of ODW surface can be found from the animation submitted with this manuscript. Addition of water droplets does not change the initiation mode, compared to the purely gaseous case in Fig. 8(a). Different from the stoichiometric mixtures, the oblique detonation is initiated through a smooth transition through a curved shock wave, for all studied flow rates. Evident from Fig. 8 is the receding transition point with increased droplet mass flow rate. Moreover, two compression waves (marked as "CWs") are observed around the transition point, which intersect with the OSW. The latter becomes concaved with respect to the flows, and hence is intensified. Due to the evaporation and motion of the droplets in the induction zone, the shocked gas temperature decreases, which is also demonstrated from the profiles at $y$ = 0.3 mm in Fig. 9. To clarify, since the data in Fig. 9 are extracted along a fixed height off the wedge (see the white lines in the insets of Fig. 8) and the OSW-ODW-CW complex moves as $f_d$ increases, the reduced pressure peaks does not mean that the intensities of the CW are weakened.

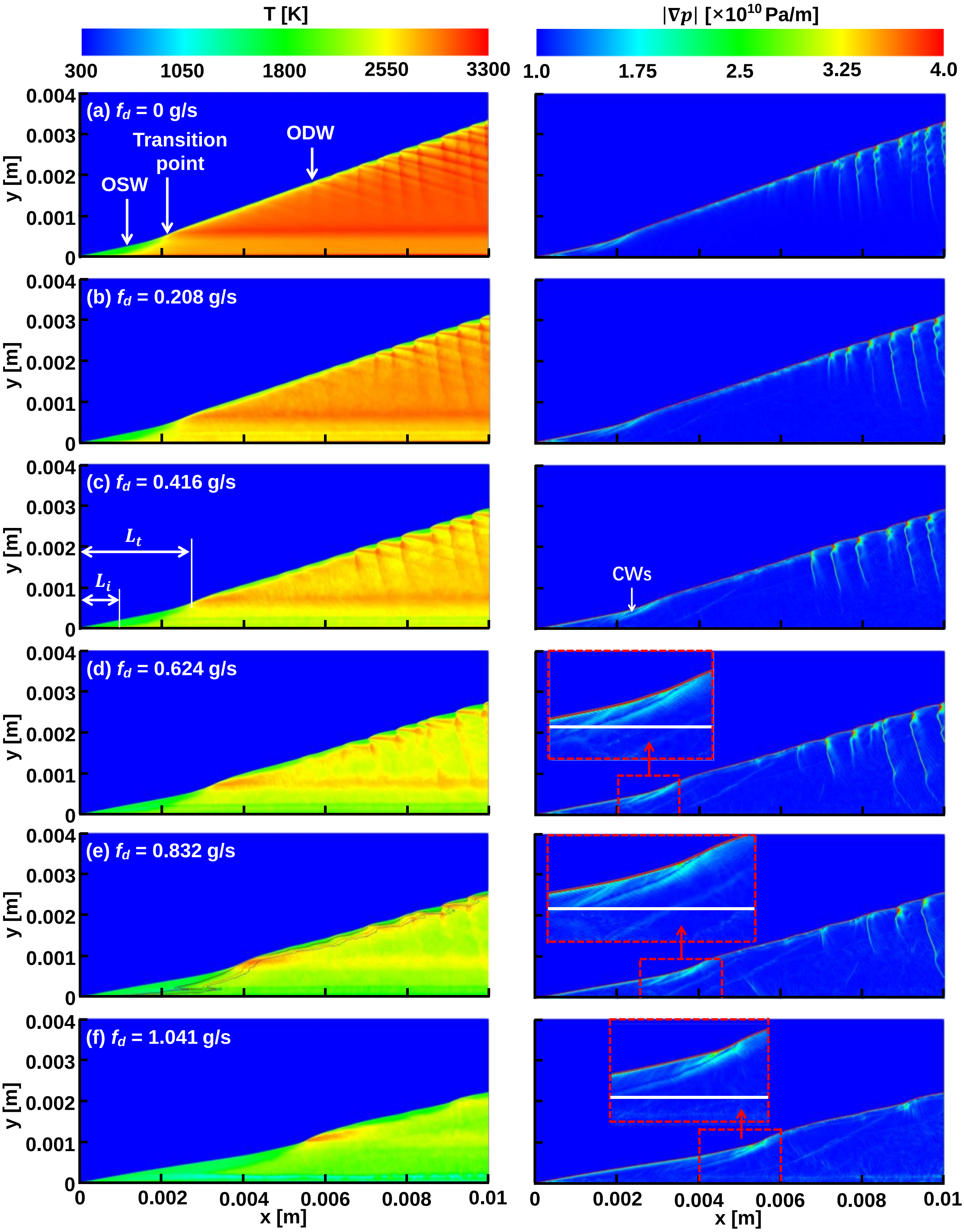

Figure 8: Gas temperature and pressure gradient magnitude from oblique detonations in lean $H_2/O_2/Ar$ mixtures with different droplet mass flow rates. Solid line in Fig 8(e): isolines of heat release rate from $3\times10^{12}$ to $5\times10^{13}$ J/(m$^3$s).

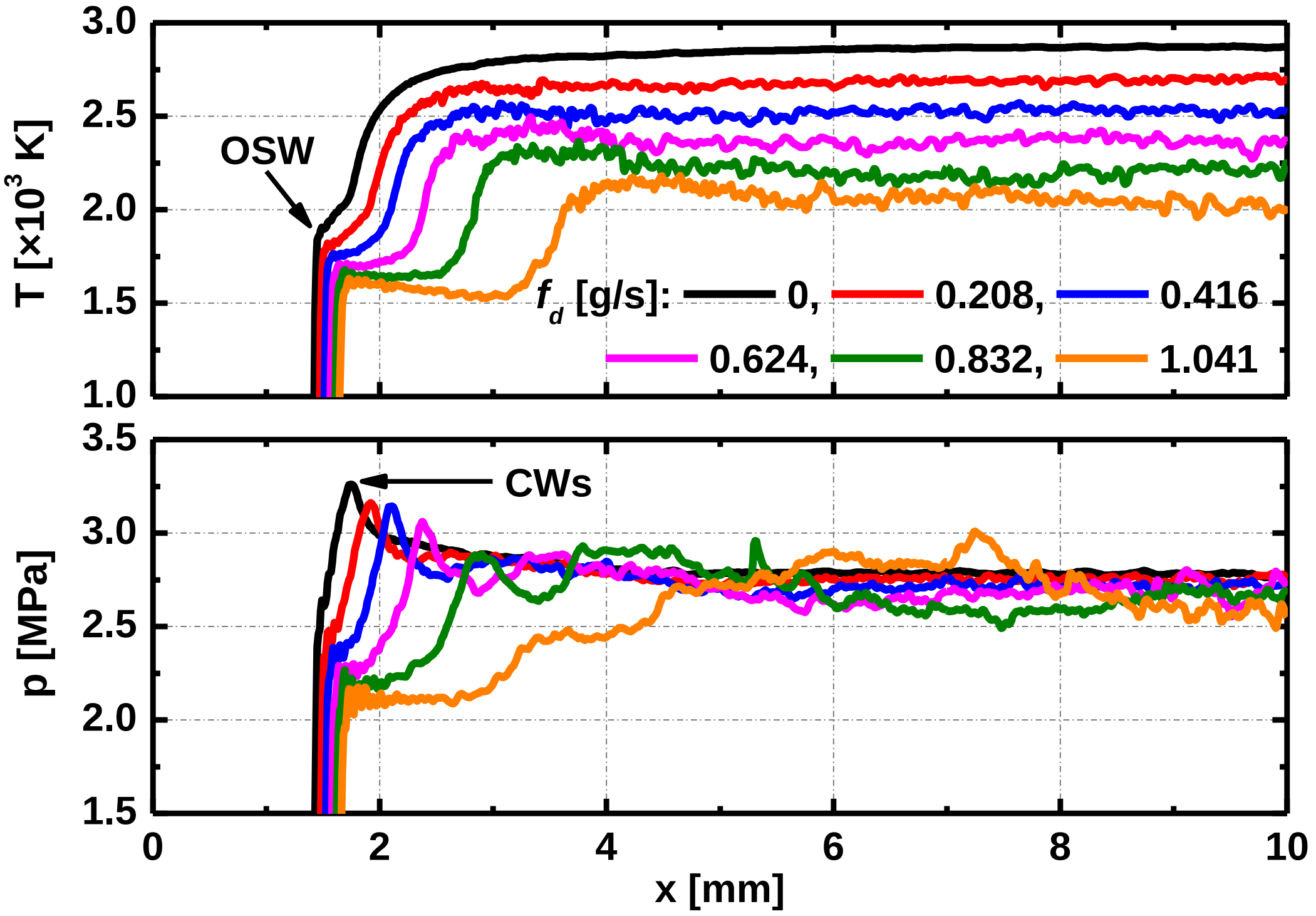

Figure 9: Profiles of gas temperature and pressure with different droplet mass flow rates. Results from $y$ = 0.3 mm. CWs: compression waves.

Figure 10 demonstrates the distributions of the combustion heat release rate with three water mass flow rates, i.e., $f_d$ = 0.208, 0.624 and 0.832 g/s. When $f_d$ is 0.208 g/s in Fig. 10(a), the deflagrative combustion is initiated at location #1 (about $x$ = 0.33 mm) near the wedge surface. The deflagration front exists when $x$ < 1.3 mm, which develops into detonation beyond $x$ = 1.3 mm, characterized by much higher heat release rate (over $1\times10^{13}$ J/m$^3$/s). This is caused by the coherent coupling between the chemical reactions and shock/compression wave (see CWs in Fig. 8). Further downstream, the reaction front smoothly evolves into the ODW. When $f_d$ is further increased to 0.624 g/s in Fig. 10(b), slight distortion of the reaction front is observed at location #2. Meanwhile, the transition from deflagration to detonation wave is also delayed, now at $x$ = 2.2 mm. In Fig. 10(c), the foregoing two features becomes more outstanding. Meanwhile, bifurcation of the reaction front appears, resulting in two initiation locations, i.e., #1 and #3, and a S-shaped morphology.

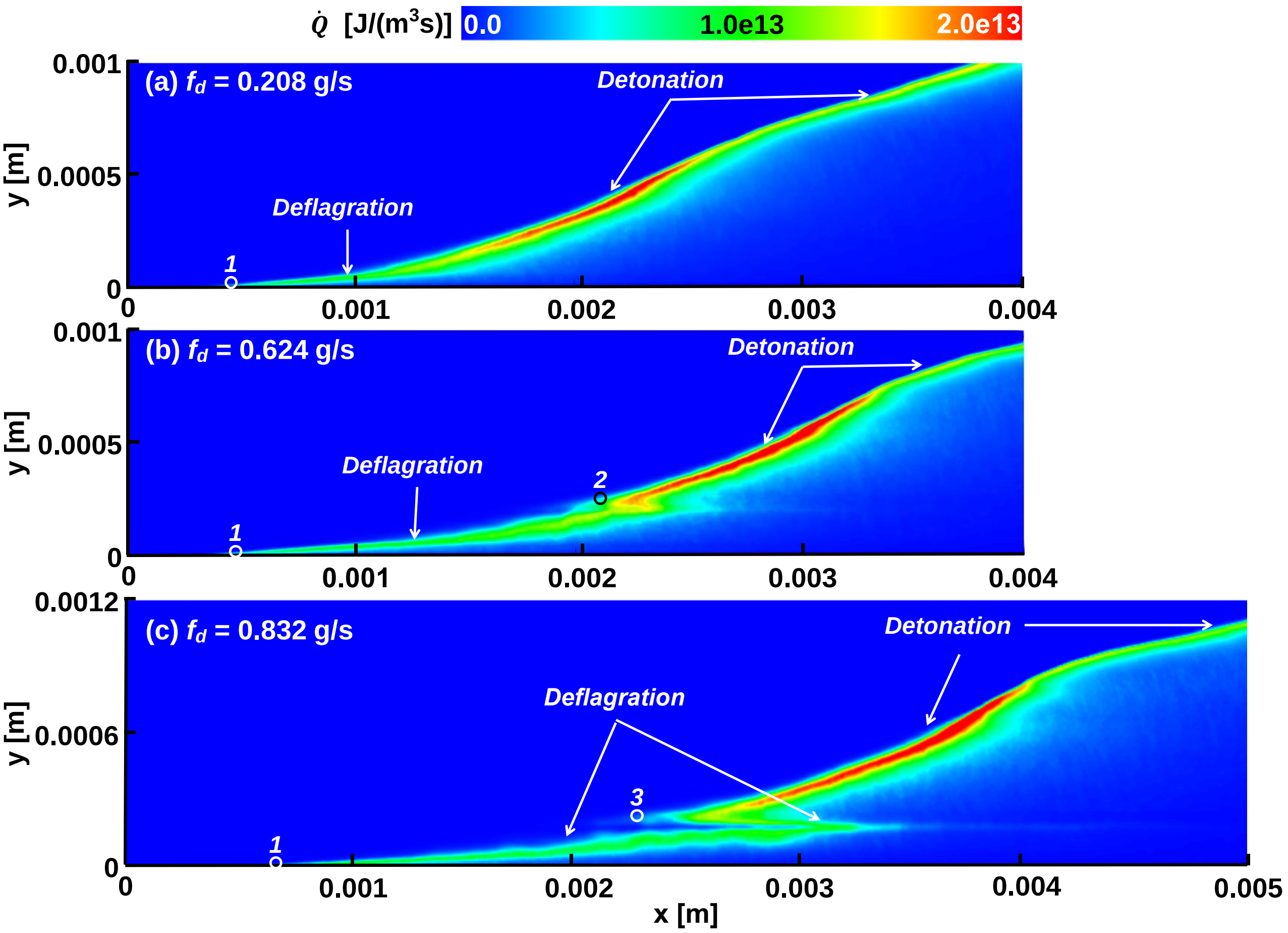

Figure 10: Distributions of combustion heat release rate with droplet mass flow rate of (a) 0.208 g/s, (b) 0.624 g/s and (c) 0.832 g/s. Circles: initiation locations.

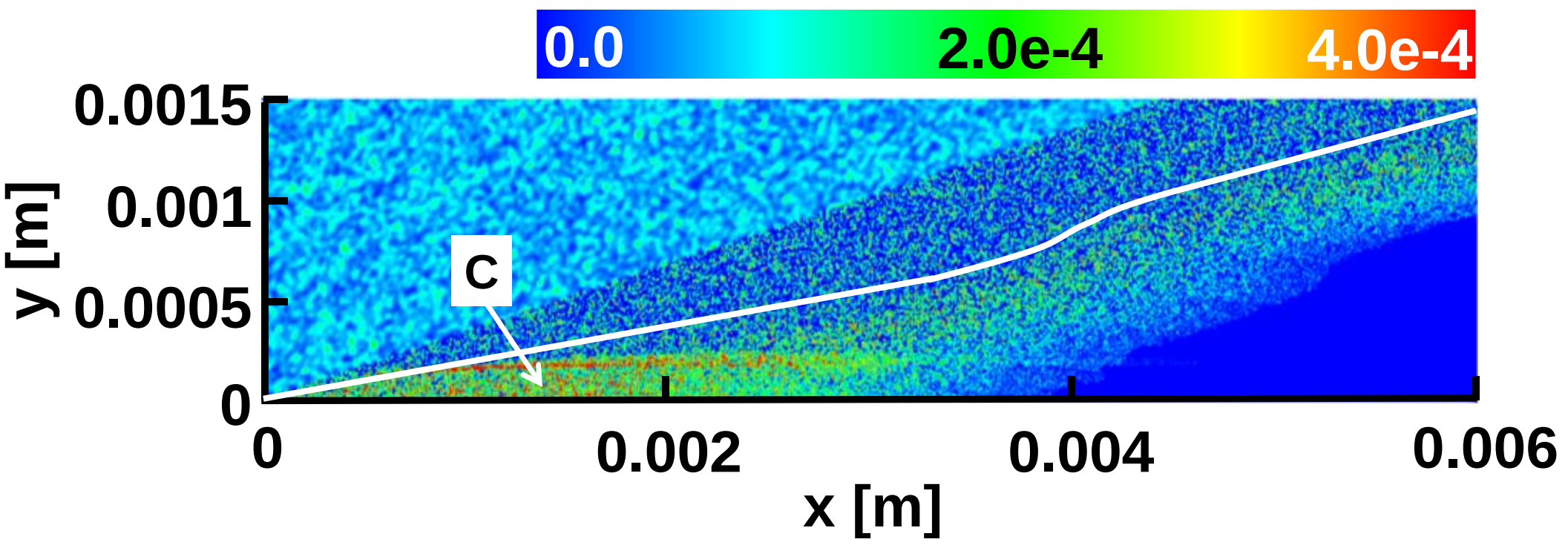

Figure 11: Distribution of droplet volume fraction. $f_d$ = 0.832 g/s. Solid line: OSW and ODW.

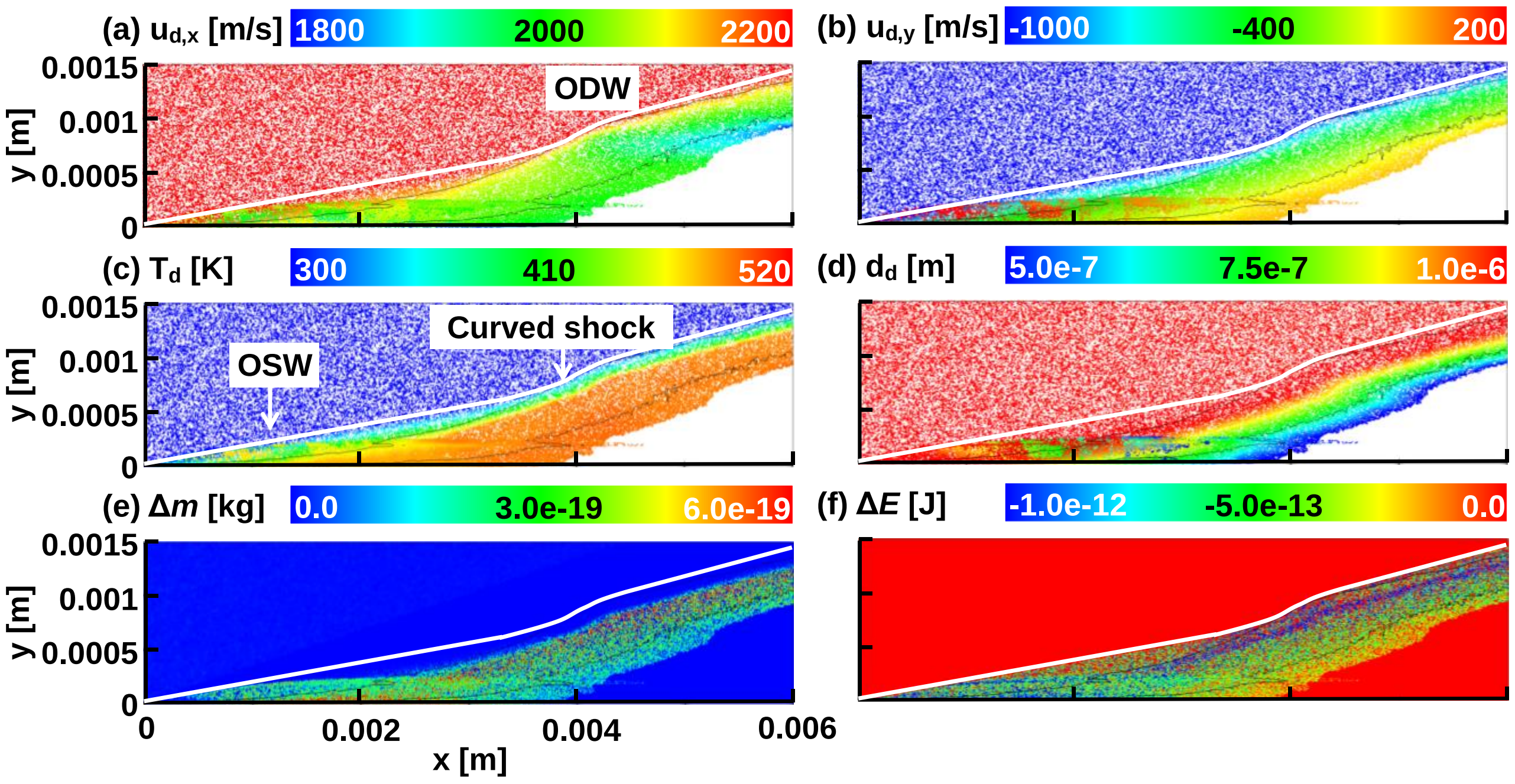

Figure 12: Lagrangian droplets colored by (a) $x$-velocity, (b) $y$-velocity, (c) temperature, (d) diameter, (e) mass transfer, and (f) energy transfer. $f_d$ = 0.832 g/s. Solid lines: heat release rate of $10^{12}$ J/(m$^3$s).

### 4.2.2 Droplet distribution

Figure 11 shows the droplet volume fraction, corresponding to the conditions in Fig. 10(c). Behind the OSW and ODW (the solid line), considerable droplets exist, featured by finite droplet volume fractions, which leads to gas-liquid heterogeneous mixtures. Relatively higher water droplet concentrations can be observed extending from the wedge edge, marked as zone C in Fig. 11. The formation of this zone share the similar mechanism with zone B in Fig. 5, due to the droplet-wall interactions. Plotted in Fig. 12 are the Lagrangian droplets colored by velocity, temperature, diameter, as well as the interphase mass and energy exchange when $f_d$ is 0.832 g/s. Behind the OSW and ODW, the droplet velocites generally have similar distributions, which is different from the situations of the abtupt transition mode in Figs. 6(a) and 6(b). This can be attributed to the flow field difference behind OSW and ODW in these two transition modes. After crossing the oblique shock wave, the droplets are heated to above 400 K, but droplet diameter remains unchanged, as shown in Figs. 12(c) and 12(d). After a certain distance behind OSW, the droplets begin to evaporate under local saturation temperature. Droplet evaporation has significant mass transfer $\Delta m$ with the surrounding gas mixtures and absorbs a lot of energy ($\Delta E < 0$) from it in Figs. 12(e) and 12(f).

Accordingly the droplet diameter decreases in Fig. 12(d). For lean $H_2/O_2/Ar$ mixtures, the droplet penetration distance normal to the detonation surface is close to that behind the OSW during thermal response time. Nonetheless, the area where droplets are present is relatively short behind the ODW due to the gas temperature behind ODW is relatively high. This can be seen from the magnitudes of mass and energy transfers in Figs. 12(e) and 12(f).

## 5. Discussion

### 5.1 Characteristic location and wave angle

Figure 13 summarizes the initiation location, transition location and wave angles from the stoichiometric and fuel-lean cases. The initiation location $L_i$ is defined as the distance from the wedge tip to the position with 10% of the maximum HRR along the wedge surface, while the transition location $L_t$ is defined as the distance ($x$-direction) from the wedge tip to the OSW-ODW transition locus. See their indications in Figs. 2 and 8. It is shown from Fig. 13(a) that for two mixtures, $L_i$ linearly increases with the droplet mass flow rate. This indicates the increased influences of droplet heating and evaporation on the initiation location. However, $L_i$ in the fuel-lean mixtures are generally smaller compared to the counterpart from the stoichiometric ones. Specifically, when $f_d$ = 0.832 g/s, for the stoichiometric and fuel-lean mixture cases, $L_i$ only increases 15.1% and 10.8% respectively, relative to the gas-only case. This is because the temperature and pressure behind OSW are generally higher in fuel-lean mixtures, for instance, see Figs. 3 and 9. The thermochemical conditions behind the OSW directly influences the reaction initiation, and therefore $L_i$ lies more upstream. Evident from Fig. 13(b) is that the transition locations from the two mixtures are more different. The transition location $L_t$ in stoichiometric mixtures changes with $f_d$ exponentially. However, only linear variation of $L_t$ is observed in fuel-lean cases. This leads to increased difference between their transition locations when the droplet loading is high, such as 0.624 and 0.832 g/s. Therefore, for fuel-lean mixtures, the sensitivity of these characteristic locations to the droplet loading variations is milder compared to that in stoichiometric cases. This implies that the fuel-lean mixtures with smooth shock-to-detonation transition mode are more stable and resilient than the stoichiometric cases. This resilence dinctinction is also observed by Iwata et al. when they consider an inhomogeneous $H_2$-air mixture [14], although no explanations are provded from their studies.

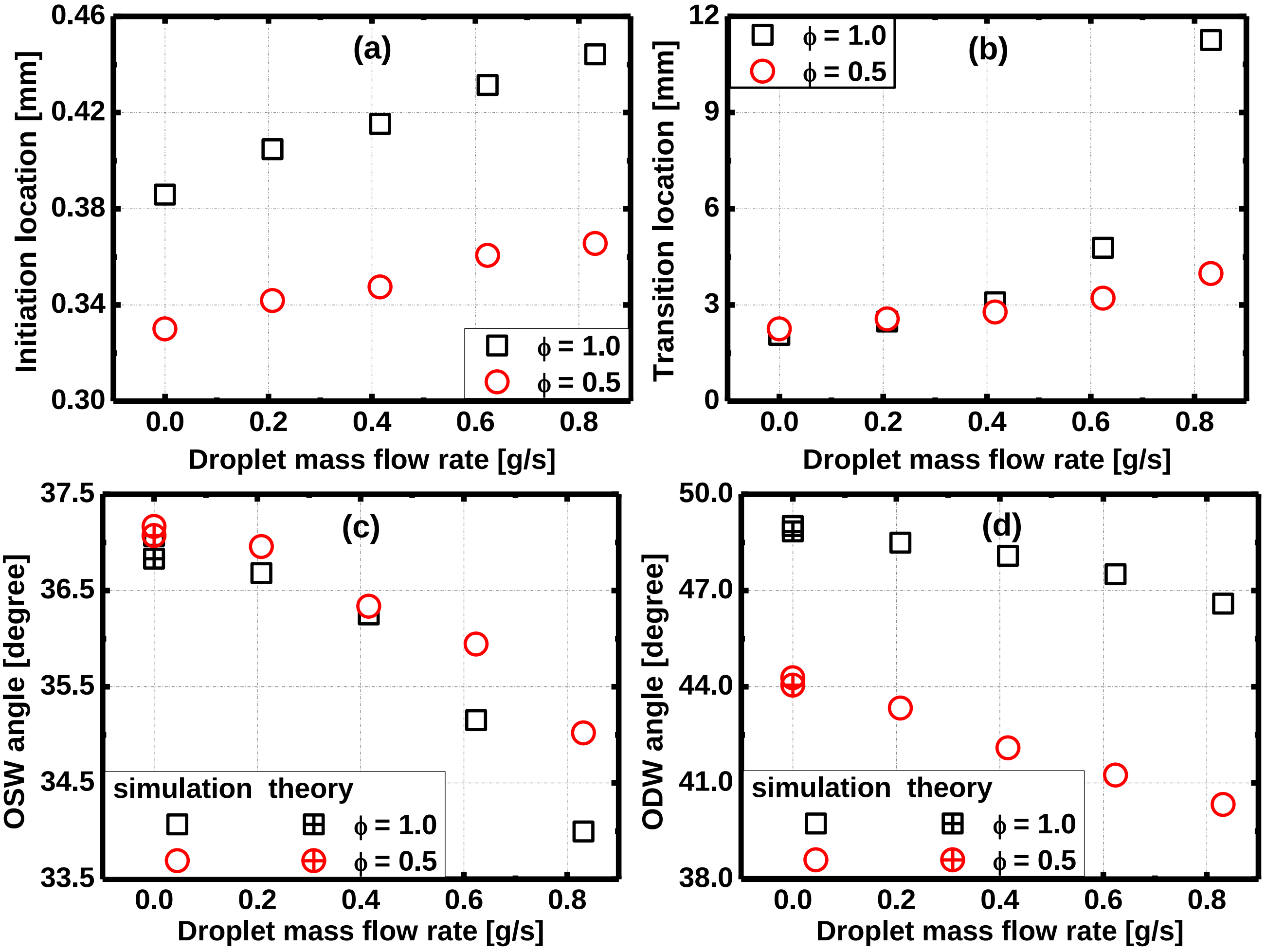

Figure 13 Initiation location (a), transition location (b), OSW angle (c), and ODW angle (d) as functions of droplet mass flow rate. Thereotical results are from shock polar analysis [55].

The evaporating water droplets also influence the OSW and ODW angles, as seen from Figs. 13(c) and 13(d). The definitions of shock/detonation angles, $\beta_{\mathrm{OSW}}$ and $\beta_{\mathrm{ODW}}$, are annotated in Fig. 1. The predicted angles in the droplet-free cases well reproduce the theoretical results from the shock polar analysis [55]. Besides, both $\beta_{\mathrm{OSW}}$ and $\beta_{\mathrm{ODW}}$ in spray detonations are lower than the values of droplet-free case. As $f_d$ increases, both angles decrease monotonically. For the OSW angle, this tendency can be justified by the fact that the oblqiue shock is attentuated by the droplet momentum and energy extraction when they cross the shock front [56,57]. This can also be comfirmed through the continuously delayed OSW position in Figs. 3 and 9. Decreased shock angle generally leads to reduced temperature and pressure behind it. This also modulates the post-shock thermodynanic conditions, thereby jointly influencing chemical reaction induction together with *in-situ* droplet heating and evaporation behind the OSW. For ODW angle, this is because higher droplet concentration would lead to more energy aborption from the gas phase and hence lower combustion heat release, which would result in smaller ODW angle based on the thermodynamic relations [7,58]. Specifically, when $f_d$ is 0.832 g/s, $\beta_{\mathrm{OSW}}$ decreases 8.28% and 5.77% respectively in the stoichiometric and fuel-lean mixtures, relative to the gas-only case. However, $\beta_{\mathrm{ODW}}$ decreases by about 4.95% and 8.93%, respectively. In spite of the above differences, the oblique shock and detonation waves have generally good stability when dispersed droplets are loaded in the oncoming flows and therefore interact with the gas mixture in the induction zone.

### 5.2 Chemical timescale and structure

The chemical explosive mode analysis [59,62] is employed to quantify the chemical reaction information of oblique detonations. It has been successfully used for analysing the induction zone of detonations [52,63,64], and detailed information about this method can be found in Refs. [59,60]. Although the smooth and abrupt OSW-ODW transition modes have been discussed in numerous studies [12,15,44,45,50], nonetheless, limited attention has been paid to their intrinsic chemical structures. Figure 14 shows the distributions of chemical timescale $\tau_{chem}$ (in logarithmic scale) of stoichiometric $H_2/O_2/Ar$ oblique detonations in droplet-free and two-phase ($f_d$ = 0.832 g/s) cases. Here, $\tau_{chem}$ is calculated from the reciprocal of the real part of eigen values $\lambda_E$ of chemical Jacobian matrix [59]. For better illustration, only chemical explosive areas with positive eigen values (hence excluding the fresh and burned areas) are shown. They correspond to the induction zones behind the shocks. The profiles of $\lambda_E$ and key thermochemical quantities along lines 1-4 in Fig. 14 are shown in Fig. 15.

We first look at the droplet-free case in Fig. 14(a). The oncoming gas changes from non-explosive to explosive state because of compression by the OSW (dotted lines in Fig. 15). Accordingly, the chemical timescale is considerably reduced, generally to less than $10^{-7.5}$ s. For instance, along line 2, $\tau_{chem}$ is continuously reduced towards the deflagration and detonation surfaces. The corresponding $\lambda_E$ gradually increases and peaks ahead of the reaction front (corresponding to $\lambda_E = 0$ [59]), as seen from Fig. 15. Near the transition location (e.g., line 1), $\tau_{chem}$ is further lowered to $10^{-8}$ s and there is a bump of $\lambda_E$, indicating the strong mixture reactivity for local detonative combustion. Beyond the transition point, the chemical timescale of the relatively narrow induction zone of ODW is generally $10^{-8}$ s.

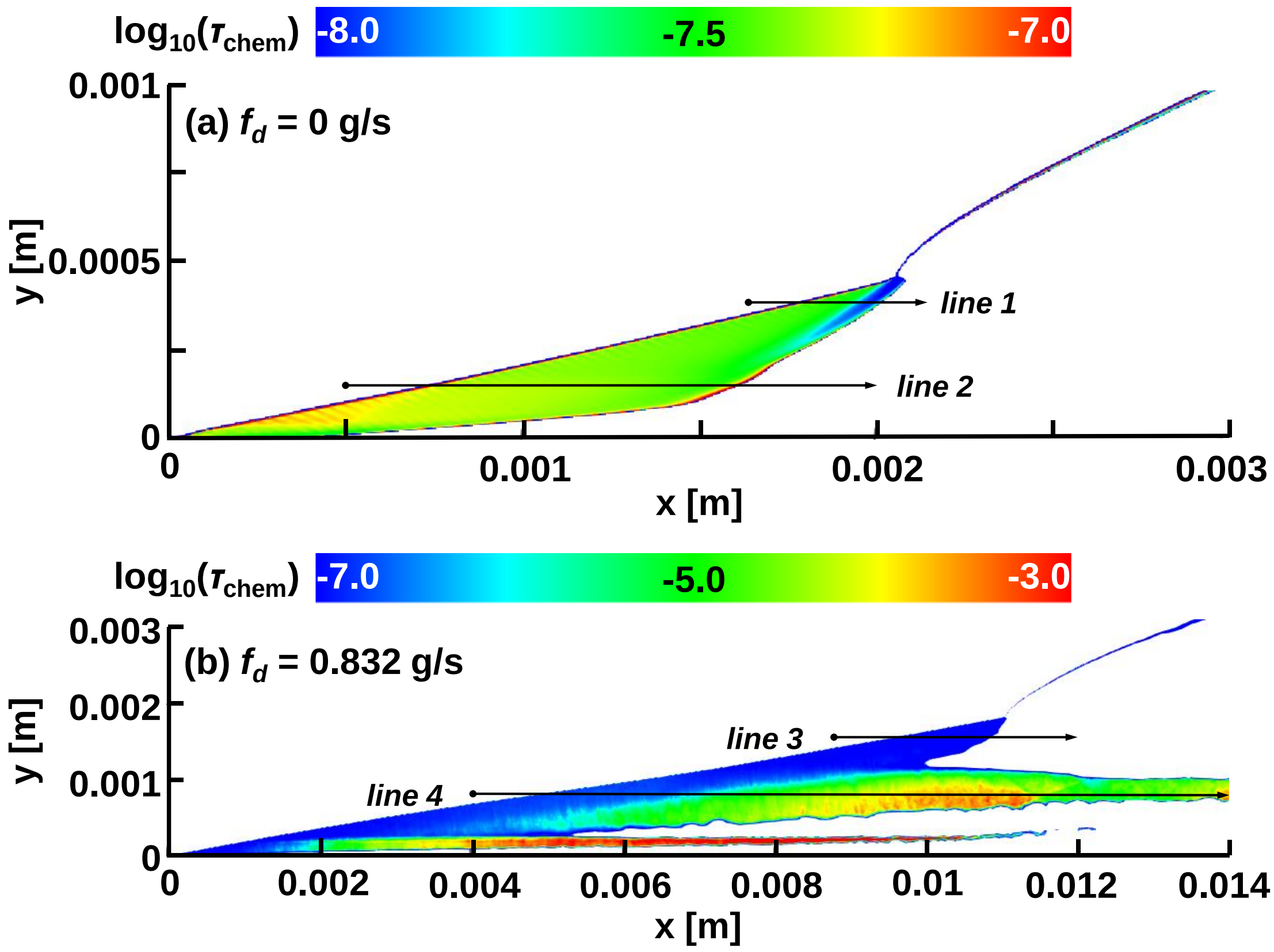

Figure 14: Distributions of chemical timescale (in second) of an oblique detonation in stoichiometric $H_2/O_2/Ar$ mixture: (a) $f_d = 0$ and (b) $f_d = 0.832$ g/s.

When the water droplets are loaded, the results in Fig. 14(b) demonstrate that the chemical timescale near the transition point increases and the chemically more explosive region (with higher $\lambda_E$) becomes narrower, compared to that of the droplet-free case. This can be found through comparing the results in Figs. 15(d) and 15(c). Moreover, there are two salient striped areas with relatively high chemical timescale (up to $10^{-3}$ s) and much lower $\lambda_E$ (e.g., see line 4 in Fig. 15d), which corresponds to zones A and B in Fig. 5. This implies that the chemical explosion propensity of the shocked gas is largely reduced due to the interactions with the evaporating water mists and hence the reaction fronts are distorted in the induction zone.

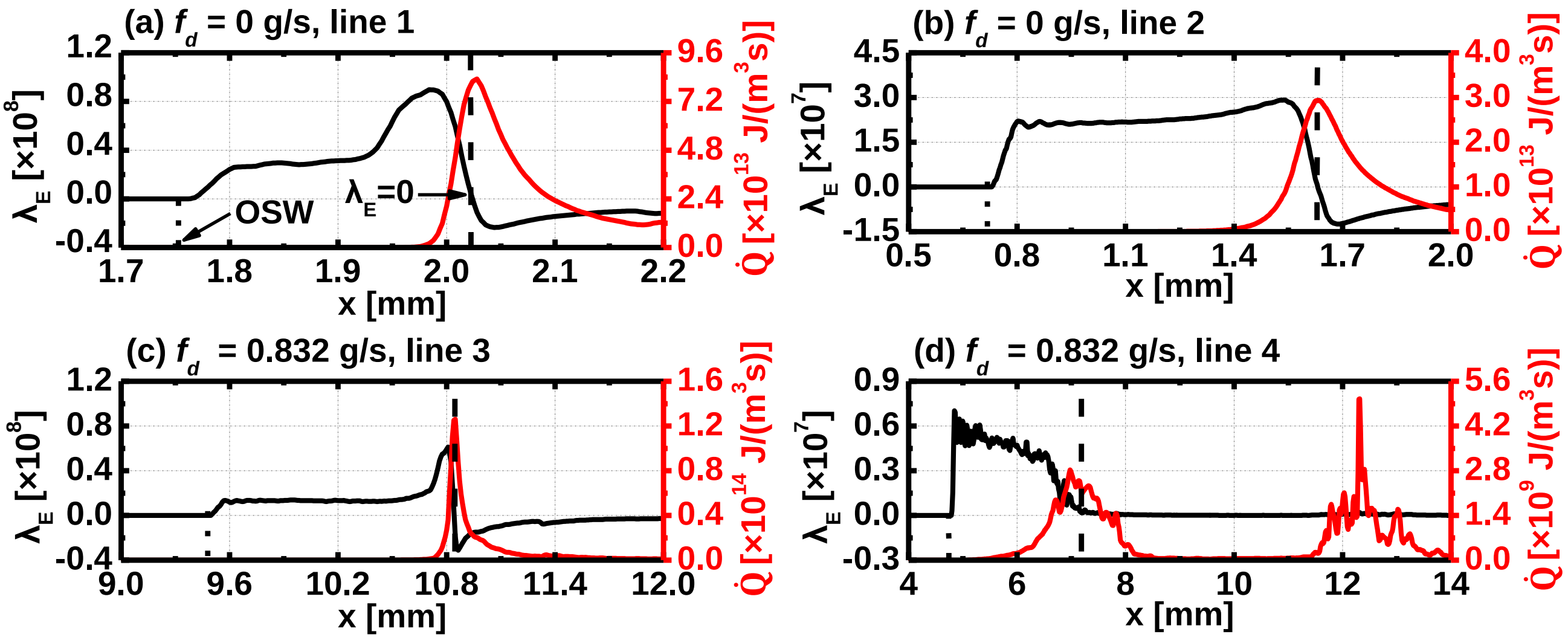

Figure 15: Profiles of chemical timescale and heat release rate at (a) line 1: $y = 3.8\times10^{-4}$ m and (b) line 2: $y = 1.5\times10^{-4}$ m for $f_d$ = 0 g/s; (c) line 3: $y = 1.55\times10^{-3}$ m and (d) line 4: $y = 8.0\times10^{-4}$ m for $f_d$ = 0.832 g/s. Stoichiometric $H_2/O_2/Ar$ mixture. Dashed line: reaction front with $\lambda_E$ = 0; dotted line: OSW.

Figures 16 and 17 show the counterpart results from the fuel-lean ($\phi$ = 0.5) $H_2/O_2/Ar$ mixture with $f_d$ = 0 and 0.832 g/s. Similar to the results in Fig. 14, the chemically explosive gas mixture also exists in the induction zones. However, we can observe interesting differences about the distribution of timescale and eigen value near the transition locations. Specifically, for smooth transition here, the gas mixture reactivity is rapidly increased behind OSW and then levels off for a distance before the deflagration front, see the variations of $\lambda_E$ in Fig. 17(a). This is caused by the compression of the curved shock, which is intensified by the compression waves. The highly reduced reaction timescale immediately behind the shock is kinetically favourable to induce the coherent interactions between the lead shock and reaction front in an ODW. Therefore, the mechanism for smooth transition is essentially different from the abrupt mode in Fig. 14 and 15. In the latter, the most chemically explosive mixture is formed by the re-compression of the shocks in the induction zone, which results in detonative combustion as demonstrated in Fig. 4. This detonation wave further intersects with the OSW and leads to an ODW.

When the water mists are added in the fuel-lean mixture, the gas reactivity near the transition point is slightly affected, see Fig. 17(c). More pronounced influence lies in zone C with higher droplet concentration (marked in Fig. 11, corresponding to line 4), and its chemical timescale is decreased to $10^{-7}$-$10^{-6.8}$ s. Even so, it is still much shorter (more reactive) than that in zones A and B in Fig. 14. Therefore, the results in Figs. 16 and 17 well justify why the ODW with smooth transition is more resilient for droplet-laden approach flows from the perspective of chemical explosiveness and characteristic timescale.

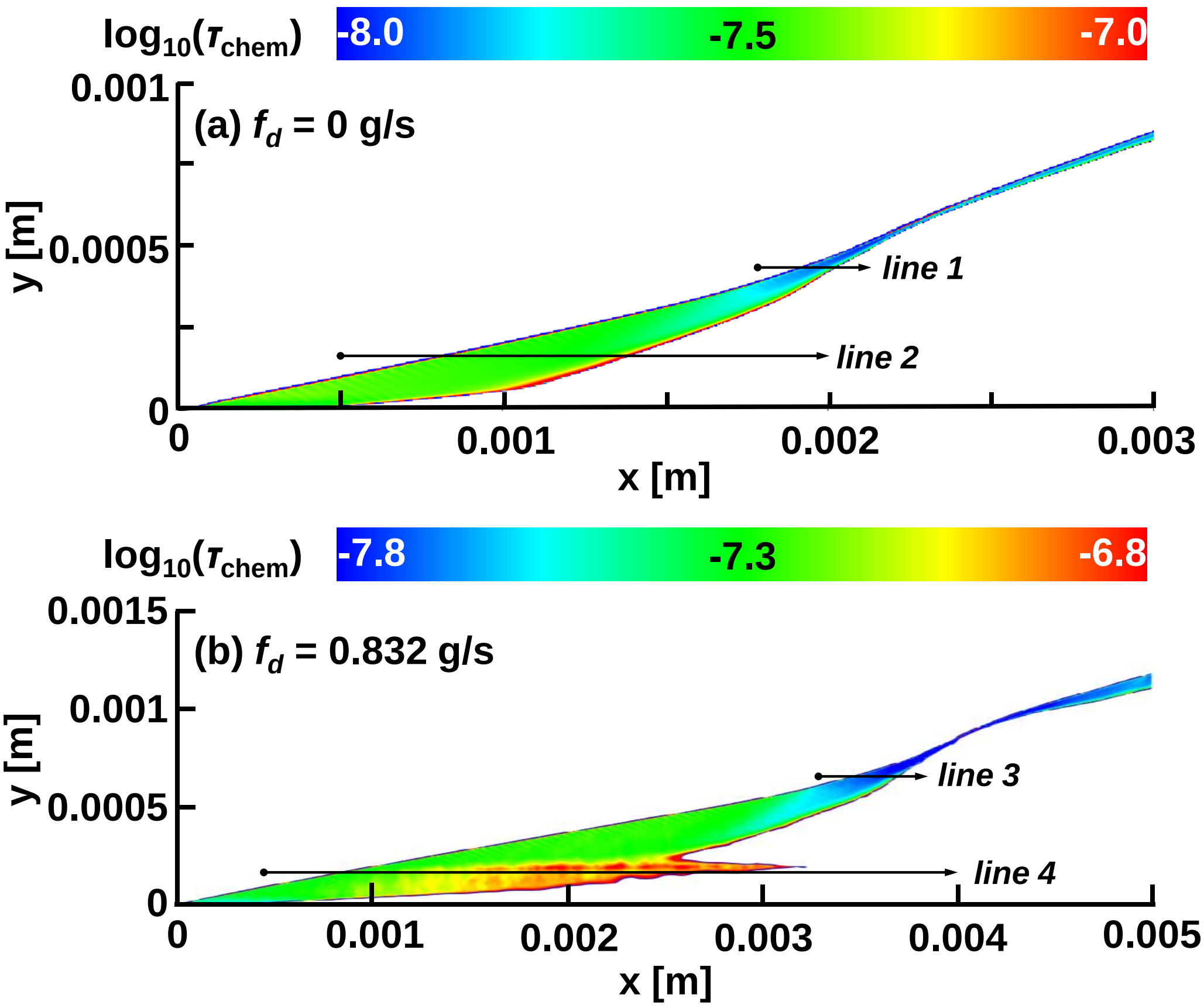

Figure 16: Distributions of chemical timescale (in second) of an oblique detonation in fuel-lean ($\phi$ = 0.5) $H_2/O_2/Ar$ mixture: (a) $f_d$ = 0 and (b) $f_d$ = 0.832 g/s.

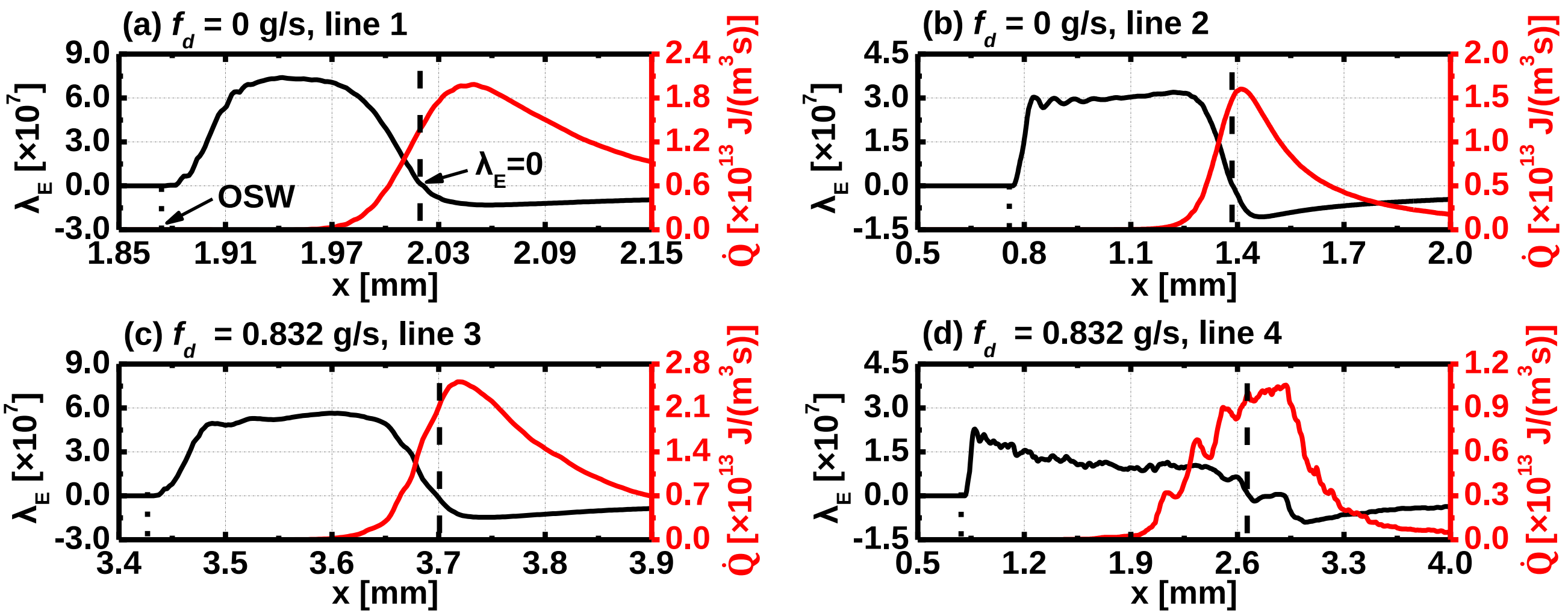

Figure 17: Profiles of chemical timescale and heat release rate at (a) line 1: $y = 4.3\times10^{-4}$ m and (b) line 2: $y = 1.6\times10^{-4}$ m for $f_d$ = 0 g/s; (c) line 3: $y = 6.5\times10^{-4}$ m and (d) line 4: $y = 1.6\times10^{-4}$ m for $f_d$ = 0.832 g/s in. Fuel-lean (ϕ = 0.5) $H_2/O_2$/Ar mixture. Dashed line: reaction front with $\lambda_E$ = 0; dotted line: OSW.

Table 1. Numerical experiments on water vapor effects on spray ODW.

| Numerical experiments | | Droplet evaporation | Chemical effects | |
|---|---|---|---|---|
| | | | As a reactant | As a third-body species |
| ϕ = 1.0 and 0.5 | a | No | No | No |
| | b | Yes | No | No |
| | c | Yes | No | Yes |
| | d | Yes | Yes | Yes |

### 5.3 Physical and chemical effects of water vapor

To elucidate how the physical and chemical effects of the water vapour, we conduct numerical experiments, following Refs. [6567]. Both stoichiometric and lean mixtures are considered and $f_d$ = 0.832 g/s. If droplet evaporation is considered, the water vapor is released from the liquid droplets. It is well known that $H_2O$ species can participate in gas phase chemical reactions, acting as reactant or third-body species. This leads to possible chemical effects of the water species in the reaction system. In our experiments, we introduce an artificial species $H_2O$(v) in the chemical mechanism, to differentiate from the water species as combustion product, $H_2O$(c). $H_2O$(v) has the same thermodynamic and transport properties as the real water species $H_2O$(c).

Four cases are compared, as tabulated in Table 1. In case a, the droplet evaporation model is turned off and hence the water vapor effects (including physical and chemical ones) are completely ruled out. Note in passing that case a is still a spray detonation modelling, considering momentum and energy transfer between dispersed and continuous phases. Case b only considers the physical effects (including compositional dilution and thermodynamic parameter modulation), whilst case c incorporates the chemical (but only as third-body species) effect. Case d has the full models, already discussed in Section 4. It is reminded that the combustion product water $H_2O$(c) plays a full role in these tests.

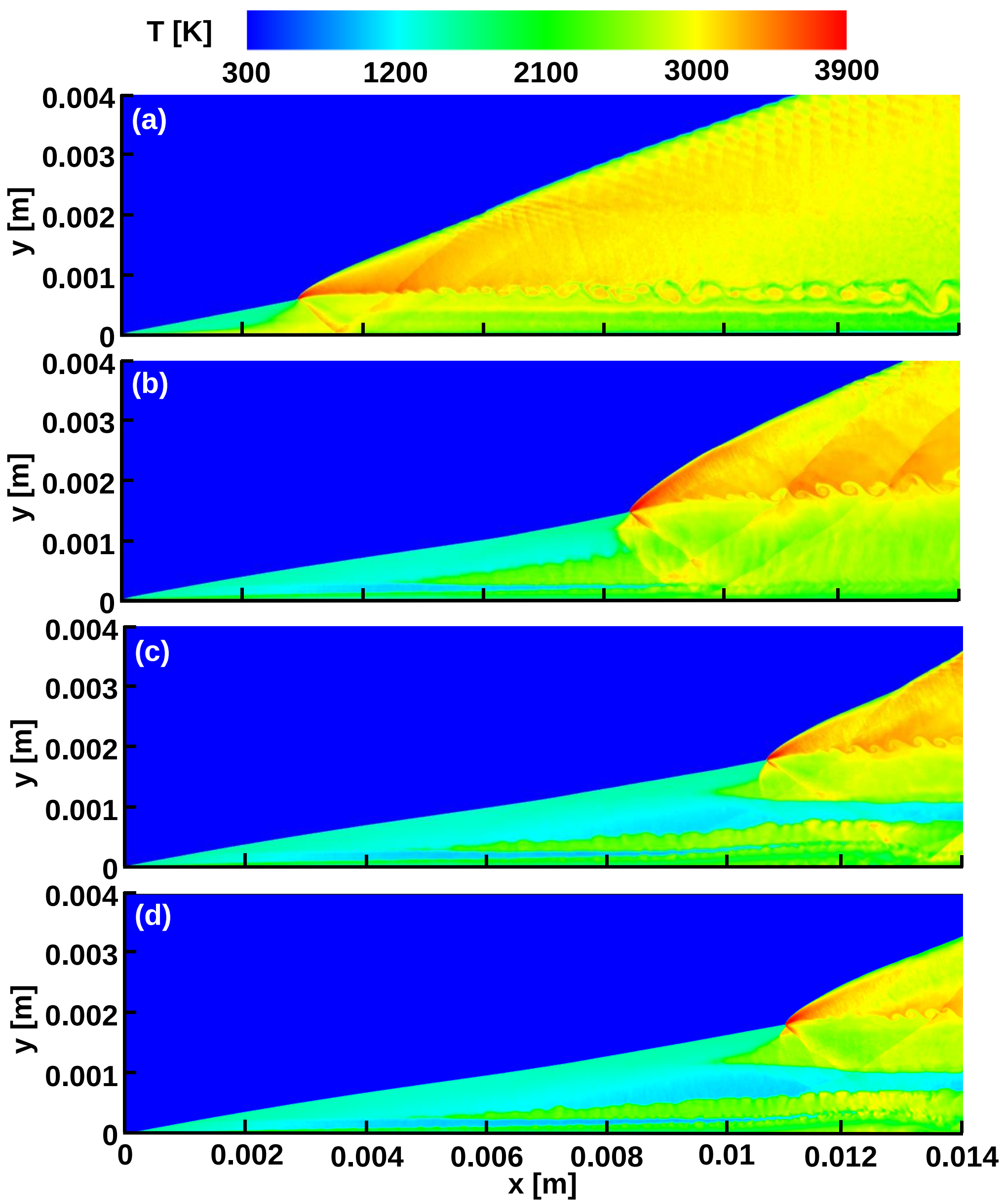

Figure 18: Gas temperature in stoichiometric $H_2/O_2/Ar$ ODWs with different roles of water vapor. $f_d$ = 0.832 g/s.

Figure 18 shows the gas temperature contours in the experiments a−d for stoichiometric

$H_2/O_2/Ar$ mixtures. For non-evaporating water droplets in case a (Fig. 18a), the structure of the induction zone is similar to the gas-only case in Fig. 2(a). Through the results in Fig. 18, one can see that the water vapor addition plays a significant role in induction zone and ODW morphology. The physical effects of water vapor can be examined in Figs. 18(a) and 18(b): addition of $H_2O$(v) considerably delays the deflagration/detonation initiation and change the induction zone structure. Figure 20 quantifies the predicted transition/initiation locations and OSW/ODW angles when different roles of $H_2O$(v) are considered. One can see that, compared to case a, transition location and ODW angle in case b increases by around 189% and 15%, respectively. Nonetheless, the initiation location and OSW angle are shown to have slight increase and decrease, respectively. This is due to the energy transfer between the surrounding gas mixtures and the post-shock water droplets, resulting in a decrease in the total energy of the post-shock gas. It means that compositional dilution and/or thermodynamic parameter variation by the water vapor results in significant differences in ODW initiation and behaviours.

When $H_2O$(v) third-body effects are considered in Fig. 18(c), the transition location further moves downstream to about $x$ = 10.7 mm and the ODW angle increases to 57.9°, compared to the results in Fig. 18(b). Meanwhile, in Fig. 18(c), a new “channel” with low temperature (also high chemical timescale as quantified in Section 5.1) appears, which disconnect the combustion waves near the multi-wave point and the wedge surface. This is due to local high concentration water vapor and hence enhanced third-body recombination elementary reactions (e.g., $H+OH+M \rightarrow H_2O+M$, $H+H+M \rightarrow H_2+M$ and $H+O_2+M \rightarrow HO_2+M$). Furthermore, the overall OSW/ODW flow field in Fig. 18(c) is close to that in case d in Fig. 18(d). Since the only difference between them is $H_2O$(v) as a reactant, we can conjecture that the $H_2O$(v) reactant effects are limited. However, a closer inspection of Figs. 18(c) and 18(d) reveals that the ODW surfaces still differ. It can be found from Fig. 20 that the ODW angle in case d is about 46.6°, much lower than that in case c. Water vapor as reactants probably shifts the elementary reactions to generate more radicals, and therefore less heat is released by exothermic reactions (which is consistent with the lower post-ODW temperature demonstrated in Fig. 18d, compared to Fig. 18c), thereby a lower ODW angle.

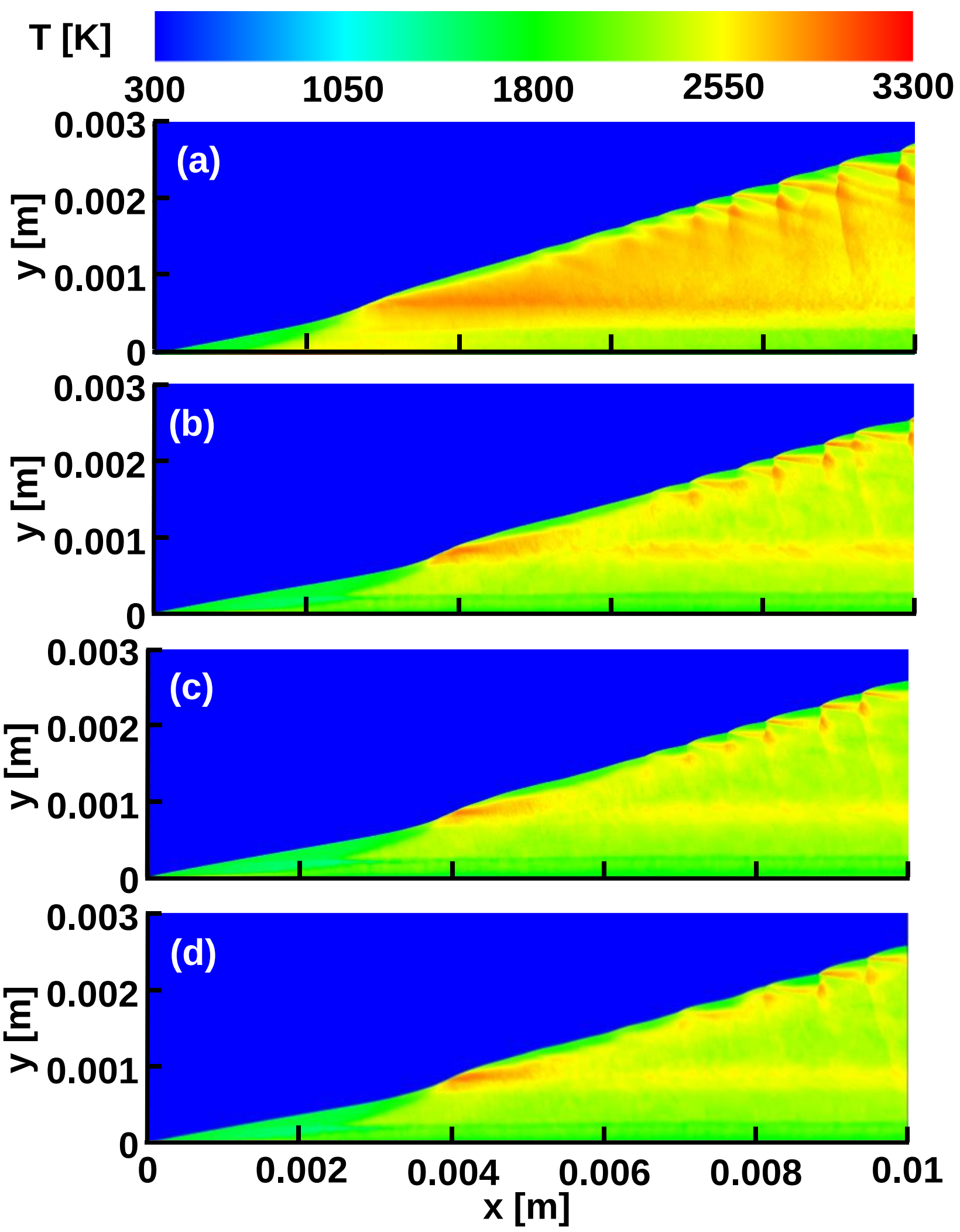

Figure 19: Gas temperature in lean $H_2/O_2/Ar$ ODWs with different roles of water vapor. $f_d$ = 0.832 g/s.

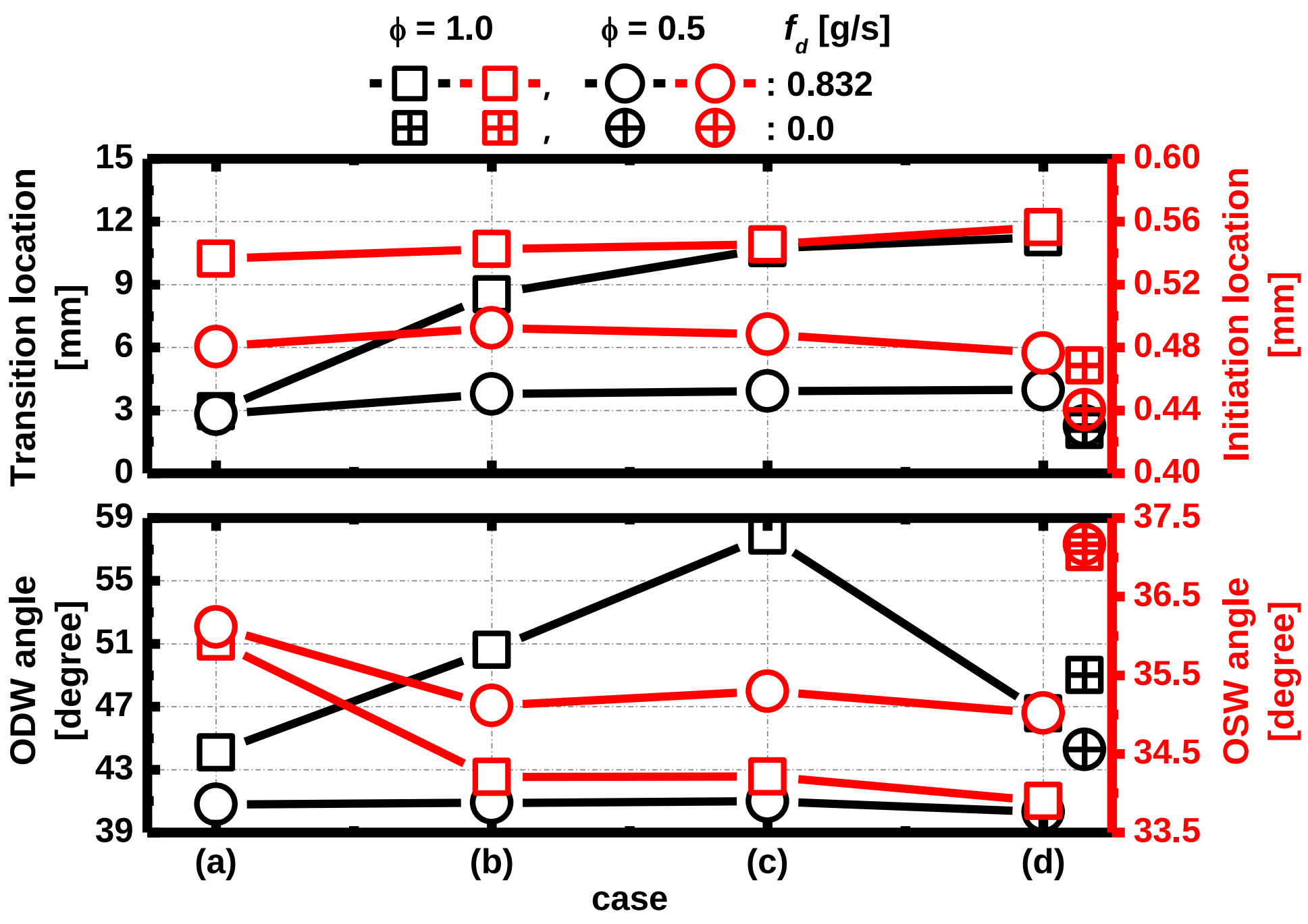

Figure 20: Transition/initiation locations and wave angles corresponding to Figs. 18 and 19.

Figure 19 show the results from the numerical experiments based on fuel-lean mixture. Through comparing Figs. 8(a) and 19(a), one can see that non-evaporating droplets slightly delay the detonation initiation. As shown in Fig. 20, the transition location increases from 2.25 to 2.81 mm. In addition, due to droplet heating and momentum absorption, the post-shock and -detonation temperatures are generally reduced. In Fig. 19(b), in which only physical effects are included, the foregoing features are more pronounced. However, different from the stoichiometric mixture results, the chemical effects of $H_2O$(v) almost do not affect the macroscopic flow structures, transition/initiation locations, and ODW angles, as quantified in Fig. 20. Besides, for both abrupt and smooth transitions, the OSW angle is negligibly affected by the chemical effects of $H_2O$(v).

To sum up, the key information from the above experiments include: (1) either physical or chemical effects of water vapor from liquid droplets do not change the OSW/ODW transition mode; (2) both effects affect the transition location and ODW angle when abrupt transition ($\phi$ = 1.0 in this study) occurs, but only the physical ones are important for smooth transition ($\phi$ = 0.5); (3) about the chemical effects of the $H_2O$(v) in abrupt transition, the third-body effects are dominant, although $H_2O$(v) also plays a discernible role as reactant for elementary reactions, thereby modulating the ODW front characteristics.

## 6. Conclusions

Two-dimensional wedge-induced oblique detonations in $H_2/O_2/Ar$ mixtures with water sprays are computationally studied with Eulerian-Lagrangian method. Stoichiometric and fuel-lean mixture compositions are considered, in which smooth and abrupt transitions from oblique shock and detonation respectively occur. The effects of water droplet mass flow rate on the change of flow and chemical structures in the induction zone, as well as the physical and chemical roles of water vapor, are discussed. The following conclusions can be drawn:

(1) The ODW can exist for the studied water droplet mass flow rates for both stoichiometric and fuel-lean mixtures. With increased droplet mass flow rate, the induction zone structure of the oblique detonation becomes complex. The deflagration front is distorted, which is caused by droplet accumulation locally and therefore higher water mist concentration, subject to different flow structures in the stoichiometric and lean mixtures. Moreover, the transition from OSW to ODW does not change when different concentrations of water sprays are loaded.

(2) For both stoichiometric and fuel-lean mixture, the initiation location and transition location monotonically increase with droplet mass flow rate. This is caused by enhanced effects of water droplet evaporation and water vapour dilution in the induction zone. In particular, the transition location increases exponentially in the stoichiometry case. Moreover, the OSW and ODW angles decrease with increasing droplet mass flow rate. In general, for fuel-lean mixtures, the sensitivity of characteristic locations to the droplet loading variations is mild. The ODW structures with smooth transition are more stable and hence more resilient than the stoichiometric cases with abrupt transition.

(3) The chemical timescale is computed with the chemical explosive method analysis. It is found that the gas between the lead shock and reaction front is intrinsically chemically explosive. For the lean mixture, the smooth transition is caused by the highly enhanced reactivity of the gas immediately behind the curved shock, intensified by the compression waves. Nonetheless, the abrupt transition results from the intersection between the detonation wave (from the coupling between shock wave and chemical reactions in the induction zone) and the OSW. With the evaporating water mists, the chemical explosion propensity of shocked gas mixtures is reduced. Besides, the degree to which the chemical timescale is reduced in fuel-lean mixtures is generally lower than that in the stoichiometric gas.

(4) Numerical experiments show that physical (composition dilution, thermodynamic state) and chemical effects (reactant or third-body species for elementary reactions) caused by water vapor from liquid droplets result in significant differences in ODW initiation and morphology. It is seen that both effects affect the transition location and ODW angle when abrupt transition occurs, but only the physical ones are important for smooth transition. As for the chemical effects of the water vapor in abrupt transition, the third-body species effects are dominant, although the role as a reactant is also discernible from the oblique detonation wave characteristics.

**Acknowledgements**

This work used the computational resources of the National Supercomputing Centre, Singapore (https://www.nscc.sg/). HG is supported by the China Scholarship Council (No. 202006680013). Professor Zhuyin Ren and Dr Wantong Wu at Tsinghua University are thanked for sharing the CEMA subroutines used in Section 5.2.

**References**

[1] K. Kailasanath, Review of propulsion applications of detonation waves, AIAA J. 38 (2015) 1698-1708.
[2] J. Urzay, Supersonic combustion in air-breathing propulsion systems for hypersonic flight, Annu. Rev. Fluid Mech. 50 (2017) 593-627.
[3] G.D. Roy, S.M. Frolov, A.A. Borisov, D.W. Netzer, Pulse detonation propulsion: Challenges, current status, and future perspective, Prog. Energy Combust. Sci. 30 (2004) 545-672.
[4] P. Wolanski, Detonative propulsion, Proc. Combust. Inst. 34 (2013) 125-158.
[5] J. Chan, J.P. Sislian, D. Alexander, Numerically simulated comparative performance of a Scramjet and Shcramjet at Mach 11, J. Propul. Power 26 (2010) 1125-1134.
[6] C. Li, K. Kailasanath, E.S. Oran, Detonation structures behind oblique shocks, Phys. Fluids 6 (1994) 1600-1611.
[7] D.T. Pratt, J.W. Humphrey, D.E. Glenn, Morphology of standing oblique detonation waves. 23rd Joint Propulsion Conference, 1987.
[8] L.F.F.E. Silva, B. Deshaies, Stabilization of an oblique detonation wave by a wedge: A parametric numerical study, Combust. Flame 121 (2000) 152-166.
[9] J. Verreault, A.J. Higgins, R.A. Stowe, Formation of transverse waves in oblique detonation, Proc. Combust. Inst. 34 (2013) 1913-1920.
[10] V.V. Vlasenko, V.A. Sabelnikov. Numerical simulation of inviscid flows with hydrogen combustion behind shock waves and in detonation waves, Combust. Explo. Shock 31 (1995) 376-389.
[11] K. Ghorbanian, J.D. Sterling, Influence of formation processes on oblique detonation wave stabilization, J. Propul. Power 12 (2015) 509-517.
[12] H. Teng, C. Tian, Y. Zhang, L. Zhou, H.D. Ng, Morphology of oblique detonation waves in a stoichiometric hydrogen-air mixture, J. Fluid Mech. 913 (2021) https://doi.org/10.1017/jfm.2020.1131.
[13] G. Fusina, J.P. Sislian, B. Parent, Formation and stability of near Chapman-Jouguet standing oblique detonation waves, AIAA J. 34 (2005) 1591-1604.
[14] K. Iwata, S. Nakaya, M. Tsue, Wedge-stabilized oblique detonation in an inhomogeneous hydrogen-air mixture, Proc. Combust. Inst. 36 (2017) 2761-2769.
[15] H. Guo, H. Yang, N. Zhao, S. Li, H. Zheng, Influence of incoming flow velocity and mixture equivalence ratio on oblique detonation characteristics, Aerosp. Sci. Technol. 118 (2021) 107088.
[16] Y. Fang, Z. Hu, H. Teng, Z. Jiang, H.D. Ng, Numerical study of inflow equivalence ratio inhomogeneity on oblique detonation formation in hydrogen-air mixtures, Aerosp. Sci. Technol. 71 (2017) 256-263.
[17] Z. Ren, B. Wang, J.X. Wen, L. Zheng,. Stabilization of wedge-induced oblique detonation waves in pre-evaporated kerosene-air mixtures with fluctuating equivalence ratios, Shock Waves 2021, https://doi.org/10.1007/s00193-021-01050-6.
[18] W. Fan, C. Yan, X. Huang, Q. Zhang, Experimental investigation on two-phase pulse detonation engine, Combust. Flame 133 (2003) 441-450.

[19] J. Kindracki, Experimental research on rotating detonation in liquid fuel-gaseous air mixtures, Aerosp. Sci. Technol. 43 (2015) 445-453.
[20] Z. Ren, B. Wang, G. Xiang, L. Zheng, Effect of the multiphase composition in a premixed fuel-air stream on wedge-induced, J. Fluid Mech. 846 (2018) 411-427.
[21] Z. Ren, B. Wang, G. Xiang, L. Zheng, Numerical analysis of wedge-induced oblique detonations in two-phase kerosene-air mixtures, Proc. Combust. Inst. 37 (2019) 3627-3635.
[22] Z. Ren, B. Wang, Q. Xie, D. Wang, Thermal auto-ignition in high-speed droplet-laden mixing layers, Fuel 191 (2017), 191(3): 176-189.
[23] H. Zhang, M. Zhao, Z. Huang, Large eddy simulation of turbulent supersonic hydrogen flames with OpenFOAM. Fuel 282 (2020) 118812.
[24] Z. Huang, M. Zhao, Y. Xu, G. Li, H. Zhang. Eulerian-Lagrangian modelling of detonative combustion in two-phase gas-droplet mixtures with OpenFOAM: Validations and verifications, Fuel 286 (2021) 119402.
[25] H.G. Weller, G. Tabor, H. Jasak, C. Fureby, A tensorial approach to computational continuum mechanics using object-oriented techniques, Comput. Phys. 12 (1998) 620-631.
[26] T. Poinsot, D. Veynante, Theoretical and numerical combustion. 3rd Edition, RT Edwards, Inc., 2012.
[27] B. E. Poling, J.M. Prausnitz, J. P. O'connell, The properties of gases and liquids. Mcgraw-hill New York, 2001.
[28] B. Mcbride, S. Gordon, M. Reno, Coefficients for calculating thermodynamic and transport properties of individual species, National Aeronautics and Space Administration, 1993.
[29] G.B. Macpherson, N. Nordin, H.G. Weller, Particle tracking in unstructured, arbitrary polyhedral meshes for use in CFD and molecular dynamics. Commun. Numer. Methods Eng. 25 (2009) 263-273.
[30] C.T. Crowe, J.D. Schwarzkopf, M. Sommerfeld, Y. Tsuji, Multiphase flows with droplets and particles. CRC Press, New York, U.S., 1998.
[31] B. Abramzon, W.A. Sirignano, Droplet vaporization model for spray combustion calculations. Int. J. Heat Mass Tran. 32 (1989) 1605-1618.
[32] M. Doble, Perry's chemical engineers' handbook. New York, U.S.A.: McGraw-Hill; 2007.
[33] W. Sutherland, LII. The viscosity of gases and molecular force, The London, Edinburgh, and Dublin Philosophical Magazine and Journal of Science 36 (2009) 507-531.
[34] E.N. Fuller, P.D. Schettler, J.C. Giddings, A new method for prediction of binary gas-phase diffusion coefficients. Ind. Eng. Chem. 58 (1966) 18-27.
[35] A.B. Liu, D. Mather, R.D. Reitz, Modeling the effects of drop drag and breakup on fuel sprays, SAE Tech. Pap. 102 (1993) 83-95.
[36] W. Ranz, W. R. Marshall, Evaporation from drops. Chem. Eng. Prog. 48 (1952) 141-146.
[37] C.T. Crowe, M.P. Sharma, D.E. Stock, The particle-source-in cell (PSI-CELL) model for gas-droplet flows, J. Fluids Eng. 99 (1977) 325-332.
[38] A. Kurganov, S. Noelle, G. Petrova, Semidiscrete central-upwind schemes for hyperbolic conservation laws and Hamilton-Jacobi equations, SIAM J. Sci. Comput. 23 (2001) 707-740.
[39] M. Ó Conaire, H.J. Curran, J.M. Simmie, W.J. Pitz, C. K. Westbrook, A comprehensive modeling study of hydrogen oxidation, Int. J. Chem. Kinet. 36 (2004) 603-622.
[40] M. Zhao, J.-M. Li, C. Teo, B.C. Khoo, H. Zhang, Effects of variable total pressures on instability and extinction of rotating detonation combustion,. Flow Turbul. Combust. 104 (2020) 261-290.
[41] M. Zhao, H. Zhang, Origin and chaotic propagation of multiple rotating detonation waves in hydrogen/air mixtures, Fuel 275 (2020) 117986.
[42] Zhao M, Zhang H. Modelling rotating detonative combustion fueled by partially pre-vaporized n-heptane sprays: Droplet size and equivalence ratio effects. Fuel 304 (2021) 121481.
[43] M. Zhao, H. Zhang, Large eddy simulation of non-reacting flow and mixing fields in a rotating detonation engine, Fuel. 280 (2020) 118534.
[44] C. Tian, H. Teng, H.D. Ng, Numerical investigation of oblique detonation structure in hydrogen-oxygen mixtures with Ar dilution. Fuel 252 (2019) 496-503.
[45] H. Teng, Z. Jiang, On the transition pattern of the oblique detonation structure, J. Fluid Mech.

713 (2012) 659-669.
[46] J.P. Sislian, H. Schirmer, R. Dudebout, J. Schumacher, Propulsive performance of hypersonic oblique detonation wave and shock-induced combustion ramjets, J. Propul. Power 17 (2012) 599-604.
[47] Y. Sheng, J.P. Sislian, Interaction of oblique shock and detonation waves, AIAA J. 21 (2012) 1008-1014.
[48] J. E. Shepherd. Shock and Detonation Toolbox - 2021 Version. California Institute of Technology
[49] H. Teng, H.D. Ng, L. Kang, C. Luo, Z. Jiang, Evolution of cellular structures on oblique detonation surfaces, Combust. Flame, 162 (2014) 470-477.
[50] H. Teng, Y. Zhang, Z. Jiang, Numerical investigation on the induction zone structure of the oblique detonation waves, Comput. Fluids 95 (2014) 127-131.
[51] J. Y. Choi, D. W. Kim, I.S. Jeung, F. Ma, V. Yang, Cell-like structure of unstable oblique detonation wave from high-resolution numerical simulation, Proc. Combust. Inst. 31 (2007) 2473-2480.
[52] Y. Xu, M. Zhao, H. Zhang, Extinction of incident hydrogen/air detonation in fine water sprays. Phys. Fluids 33 (2021) 116109.
[53] H. Watanabe, A. Matsuo, A. Chinnayya, K. Matsuoka, A. Kawasaki, J. Kasahara, Numerical analysis of the mean structure of gaseous detonation with dilute water spray, J. Fluid Mech. 887 (2020), A4. https://doi.org/10.1017/jfm.2019.1018.
[54] H. Teng, H. D. Ng, Z. Jiang, Initiation characteristics of wedge-induced oblique detonation waves in a stoichiometric hydrogen-air mixture, Proc. Combust, Inst. 36 (2017) 2735-2742.
[55] H. Guo, N. Zhao, H. Yang, S. Li, H. Zheng, Analysis on stationary window of oblique detonation wave in methane-air mixture. Aerosp. Sci. Technol. 118 (2021) 107038.
[56] Z. Huang, H. Zhang, On the interactions between a propagating shock wave and evaporating water droplets. Phys. Fluids 32 (2020) 123315.
[57]J. Kersey, E. Loth, D. Lankford, Effect of evaporating droplets on shock waves. AIAA J. 48 (2010) 1975-1985.
[58] S. A. Ashford, G. Emanuel, Wave angle for oblique detonation waves. Shock Waves 3 (1994) 327-329.
[59] T.F. Lu, C.S. Yoo, J.H. Chen, C.K. Law, Three-dimensional direct numerical simulation of a turbulent lifted hydrogen jet flame in heated coflow: A chemical explosive mode analysis. J. Fluid Mech. 652 (2010) 45-64.
[60] D.A. Goussis, I. Hong, H.N. Najm, S. Paolucci, M. Valorani, The origin of CEMA and its relation to CSP. Combust. Flame, 227 (2021) 396-401.
[61] W. Xie, W. Wu, Z. Ren, H. Liu, M. Ihme, Effects of evaporation on chemical reactions in counterflow spray flames. Phys. Fluids 33 (2021) 065115.
[62] S.H. Lam, D.A. Goussis, The CSP method for simplifying kinetics. Int. J. Chem. Kinet. 26 (1994) 461-486.
[63] T. Jaravel, O. Dounia, Q. Malé, O Vermorel, Deflagration to detonation transition in fast flames and tracking with chemical explosive mode analysis. Proc. Combust. Inst. 38 (2020) 3529-3536.
[64] M. Zhao, Z. Ren, H. Zhang, Pulsating detonative combustion in n-heptane/air mixtures under off-stoichiometric conditions. Combust. Flame, 226 (2021) 285-301.
[65] W. Zhang, X. Gou, Z. Chen, Effects of water vapor dilution on the minimum ignition energy of methane, n-butane and n-decane at normal and reduced pressures. Fuel 187 (2017) 111-116.
[66] F. Liu, H. Guo, G.J. Smallwood, The chemical effect of $CO_2$ replacement of $N_2$ in air on the burning velocity of $CH_4$ and $H_2$ premixed flames. Combust. Flame 133 (2003) 495–497.
[67] Z. Yu, H. Zhang, Reaction front development from ignition spots in n-heptane/air mixtures: low-temperature chemistry effects induced by ultrafine water droplet evaporation. Phys. Fluids 33 (2021) 083312.